\documentstyle[emulateapj]{article}
\input epsf

\def\calk{{\cal K}}
\def\cale{{\cal E}}

\def\etal{{\rm et~al. }}
\def\spose#1{\hbox to 0pt{#1\hss}}
\def\simlt{\mathrel{\spose{\lower 3pt\hbox{$\mathchar"218$}}
     \raise 2.0pt\hbox{$\mathchar"13C$}}}
\def\simgt{\mathrel{\spose{\lower 3pt\hbox{$\mathchar"218$}}
     \raise 2.0pt\hbox{$\mathchar"13E$}}}
\def\beq{\begin{equation}}
\def\eeq{\end{equation}}
\def\bce{\begin{center}}
\def\ece{\end{center}}
\def\bea{\begin{eqnarray}}
\def\eea{\end{eqnarray}}
\def\ben{\begin{enumerate}}
\def\een{\end{enumerate}}

\def\brr{\begin{array}}
\def\err{\end{array}}

\def\etal{{\rm et~al. }}

\def\nh1{n_{\rm HI}}

\def\p1dk{P_{\rm 1D}(k)}
\def\simlt{\mathrel{\spose{\lower 3pt\hbox{$\mathchar"218$}}
     \raise 2.0pt\hbox{$\mathchar"13C$}}}

\newcommand{\xibar}{\overline{\xi}}
\newcommand{\zbar}{\overline{z}}

\newcommand{\deltab}{\overline{\delta}}

\def\Or{{\cal O}}

\newcommand{\vtheta}{\vec{\theta}}

\newbox\grsign \setbox\grsign=\hbox{$>$} \newdimen\grdimen \grdimen=\ht\grsign
\newbox\simlessbox \newbox\simgreatbox
\setbox\simgreatbox=\hbox{\raise.5ex\hbox{$>$}\llap
     {\lower.5ex\hbox{$\sim$}}}\ht1=\grdimen\dp1=0pt
\setbox\simlessbox=\hbox{\raise.5ex\hbox{$<$}\llap
     {\lower.5ex\hbox{$\sim$}}}\ht2=\grdimen\dp2=0pt

\received{2002 October 16}
\begin{document} 

\title{Correlation between galaxies and QSO in the SDSS:\\
a signal from gravitational lensing magnification?}

\author{Enrique Gazta\~naga}
\affil{Institut d'Estudis Espacials de Catalunya, IEEC/CSIC,
Gran Capit\'a 2-4, 08034 Barcelona, Spain}



\begin{abstract}
  
  We report a detection of galaxy-QSO cross-correlation $w_{GQ}$ in the Sloan
  Digital Sky Survey (SDSS) Early Data Release (EDR) over $0.2-30$ arc-minute
  scales.  We cross-correlate galaxy samples of different mean depths
  $r'=19-22$ ($\zbar_G =0.15-0.35$) with the main QSO population ($i'_Q
  <19.1$) at $\zbar_Q \simeq 1.6$.  We find significant positive correlation in all cases
except for the faintest QSOs, as expected if the signal were due to weak lensing magnification.
  The amplitude of the signal on arc-minute scales is about $20\%$ at $\zbar_G
  =0.15$ decreasing to $10\%$ at $\zbar_G =0.35$. This is a few times larger
  than currently expected from weak lensing in the $\Lambda$CDM models 
  but confirms, at a higher significance, previous
  measurements by several groups.  When compared to the
  galaxy-galaxy correlation $w_{GG}$, a  weak lensing interpretation
 indicates a strong and
  steep non-linear amplitude for the underlaying matter fluctuations: $\sigma
  \simeq 400$ on scales of $0.2$ Mpc/h, in contradiction with non-linear
  modeling of $\Lambda$CDM fluctuations. We also detect a normalized skewness
  (galaxy-galaxy-QSO correlation) of $S_3 \simeq 21 \pm 6$ at ${\zbar} \simeq
  0.15$ ($S_3 \simeq 14 \pm 4$ at ${\zbar} \simeq 0.35$), which several sigma low
as compared to standard $\Lambda$CDM expectations. These
  observational trends can be reconciled with lensing in a flat $\Lambda$ universe with
  $\sigma_8 \simeq 1$, provided the linear spectrum is steeper ($n \simeq 1$)
  than in the $\Lambda$CDM model on small (cluster) scales. Under this interpretation, 
  the galaxy distribution traces the
  matter variance with an amplitude that is 100 times smaller: ie galaxies
  are anti-bias with $b\simeq 0.1$ on small scales, increasing to $b\simeq 1 $ at $\simeq
  10$ Mpc/h.

\end{abstract}


\section{Introduction}

Weak gravitational lensing by foreground large-scale matter density
fluctuations could introduce significant density variations in flux-limited
samples of high redshift objects, such as QSOs. This is sometimes called
magnification bias or cosmic magnification. In principle, it is possible to
separate the intrinsic density fluctuations in QSO samples from the weak
lensing magnification signal by cross-correlating the QSOs with a low redshift
galaxy sample. This allows a direct measurement of how the galaxy distribution
traces the underlaying mass distribution.  Much work have been done in this
direction both on theory and observations (for recent reviews see Norman \&
Williams 2000, Bartelmann \& Schneider 2001, Benitez \etal 2001, Guimar\~aes,
van de Bruck and Brandenberger 2001).

Correlation of low redshift galaxies and high-redshift AGNs or QSOs typically
find significant excesses of foreground objects around the QSO positions
(Seldner \& Peebles 1979,
Tyson 1986; Fugmann 1988,1990; Hammer \& Le F\'{e}vre 1990; Hintzen \etal
1991; Drinkwater \etal 1992; Thomas \etal 1995; Bartelmann \& Schneider
1993, 1994; Bartsch, Schneider, \& Bartelmann 1997; Seitz \& Schneider;
Ben\'{\i}tez \etal 1995; Ben\'{\i}tez \& Mart\'{\i}nez-Gonz\'{a}lez 1995,
1997; Ben\'{\i}tez, Mart\'{\i}nez-Gonz\'{a}lez \& Mart\'{\i}n-Mirones
1997;Norman \& Impey 1999;Norman \& Williams 2000). See also
 Burbidge \etal 1990 (and references therein) for individual associations
and different statistical indications.  On the other hand, 
there has also been reports of significant anti-correlation between 
galaxy groups and faint QSOs (see Croom \& Shanks 1999 and references
therein) which could also be explained by weak lensing magnification because
of the flater slope of the faint QSO number counts.
Even though the shape
of the galaxy-QSO cross-correlation has not yet been well constrained, these
results seem qualitatively in agreement with the magnification bias effect,
but the amplitude of the correlation is found to be higher than that expected
from gravitational lensing models based on $\Lambda$CDM. 

What is the origin of this discrepancy? Part of the problem could be due to
the lack of well-defined and homogeneous samples. After all we are looking for
a small effect, may be as low as $1\%$, and any systematics in the sample
definition is likely to introduce cross-correlations at this level (see
\S 2 below). On the other hand we know little about clustering
of dark matter  on sub-megaparsec scales. Could the discrepancy be
due to real deviations from the  $\Lambda$CDM paradigm?

Recently, M\'enard \& Bartelmann (2002) and M\'enard, Bartelmann \& Mellier
(2002) have explored the interest of the SDSS to cross-correlate foreground
galaxies with background QSO.  The SDSS collaboration made an early data
release (EDR) publicly available on June 2001.  The EDR includes around a
million galaxies and 4000 QSOs distributed within a narrow strip of 2.5
degrees across the equator (see Stoughton et al 2002 for details).  As the
strip crosses the galactic plane, the data is divided into two separate sets
in the North and South Galactic caps. The SDSS collaboration has presented a
series of analysis (Zehavi \etal 2002, Blanton \etal 2002, Scranton et a.l
2002, Connolly \etal 2002, Dodelson \etal 2002, Tegmark \etal 2002, Szalay
\etal 2002, Szapudi 2002) of large scale angular clustering on the North
Galactic strip, which contains data with the best seeing conditions in the
EDR.  Gazta\~naga (2002a, 2002b) presented a first study of bright ($g' \simeq
20$) SDSS galaxies in both the South and North Galactic EDR strip, centering
the analysis on the comparison of clustering to the APM galaxy Survey (Maddox
\etal 1990).

In this paper we will follow closely M\'enard, Bartelmann \& Mellier (2002,
MBM02 from now on) proposal to study the galaxy-QSO cross-correlation signal
in the EDR/SDSS.  The paper is organized as follows.  In section \S2 we
present the samples used and the galaxy and QSO selection.  Section \S3 is
dedicated to the extinction contamination. Section \S4 and \S5 presents the
main results and its interpretation, while \S6 and \S7 are dedicated to
a discussion and a listing of conclusions.

\section{QSO and galaxy Samples}

The galaxy samples are obtained from the EDR and converted into pixel maps of
different resolutions as described in Gazta\~naga (2002a,2002b).  
 We select
objects from an equatorial SGC (South Galactic Cap) strip 2.5 wide ($-1.25 <
DEC <1.25$ degrees.)  and 66 deg. long ($351 < RA < 56 $ deg.), which will be
called EDR/S, and also from a similar NGC (North Galactic Cap) 2.5 wide and 91
deg.  long ($145 < RA < 236 $ deg.), which will be called EDR/N.  These strips
(SDSS numbers 82N/82S and 10N/10S) correspond to some of the first runs of the
early commissioning data (runs 94/125 and 752/756) and have variable seeing
conditions.  Runs 752 and 125 are the worst with regions where the seeing
fluctuates above 2''.  Runs 756 and 94 are better, but still have seeing
fluctuations of a few tenths of arc-second within scales of a few
degrees.  These seeing conditions could introduce
large scale gradients because of the corresponding variations in the
photometric reduction (eg star-galaxy separation) that could manifest as large
scale number density gradients (see Scranton et al 2001 for a detail account
of these effects).  We will test our results against the possible effects of
seeing variations, by  using a seeing mask (see \S 3.1).

Redshift
targets in the SDSS are selected in $r'$ for galaxies and $i'$ for QSO. Here
we use $i'$ for QSO and both $i'$ and $r'$ for galaxies.  
\footnote{We will use $z',i',r',g',u'$ for 'raw', uncorrected magnitudes, and
  $z^*,i^*,r^*,g^*,u^*$ for extinction corrected magnitudes. For example,
  according to Schlegel \etal (1998) $r'=18$ corresponds roughly to an average
  extinction corrected $r^* \simeq 17.9$ for a mean differential extinction
  $E(B-V) \simeq 0.03$.}
The first choice has the interest that both galaxies
and QSO come from the same photometric reduction and are therefore subject to
similar systematics. Galactic extinction is also smaller in this band. The
second choice provides a comparison with previous results on galaxy
clustering. To maximize the number of galaxies, and therefore the possibility
of lensing, we choose broad magnitude bands. We will focus our results in
comparing a nearby sample $r'=19-17.5$ ($\zbar \simeq 0.15$) with a distant one
$i', r'=22-17.5$ ($\zbar \simeq 0.35$).  Galaxies selected with $i'<22$ are
almost the same (within few percent) to galaxies in $r'<22$ as the
K-correction cancels out with the color evolution (see Fukugita \etal 1996).
Here we take the $i'<22$ band as our nominal choice to minimize extinction
(this will give larger area coverage, see below). As we will show, both
galaxy-galaxy and galaxy-QSO cross-correlations turn out to be almost
identical in $i'<22$ and $r'<22$, the only difference being slightly smaller
errors in the $i'<22$ sample.  For the bright sample we stick to $r'<19$ to
provide a more direct comparison with previous results on galaxy-galaxy
clustering. Using the redshift distributions in Dodelson \etal (2002),
the mean redshift for $17.5<r'<19$ (with $\simeq 81000/117000$ galaxies)
is $\zbar \simeq 0.15$, while  $17.5< i',r' <22$ (with $\simeq 480000 $
galaxies within our mask)
have   $\zbar \simeq 0.35$. We will center our analysis over
these two samples, which 
will be sometimes refered to as $r'<19$ and $i'<22$.

\begin{figure*} 
\centering{
{\epsfxsize=18cm \epsfbox{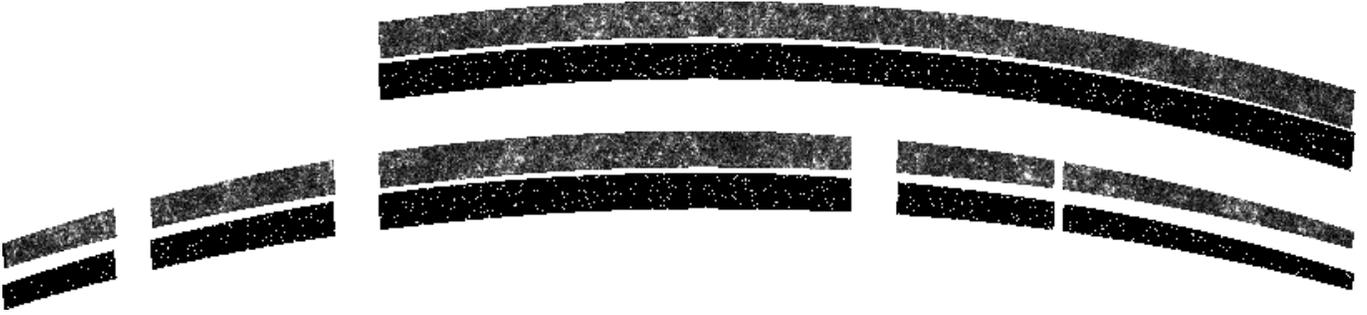}}
}
\caption[Fig2] {\label{pixelMap} 
Equatorial projections galaxy and QSO maps. 
The top pair of slices correspond to
galaxy  and QSOs in SDSS EDR/S ($2.5 \times 60$ sqr.deg).
The bottom two slices shows galaxies and QSOs in SDSS EDR/N
($2.5 \times 90$ sqr.deg).}
\end{figure*}

\begin{figure*} 
\centering{
{\epsfxsize=18cm \epsfbox{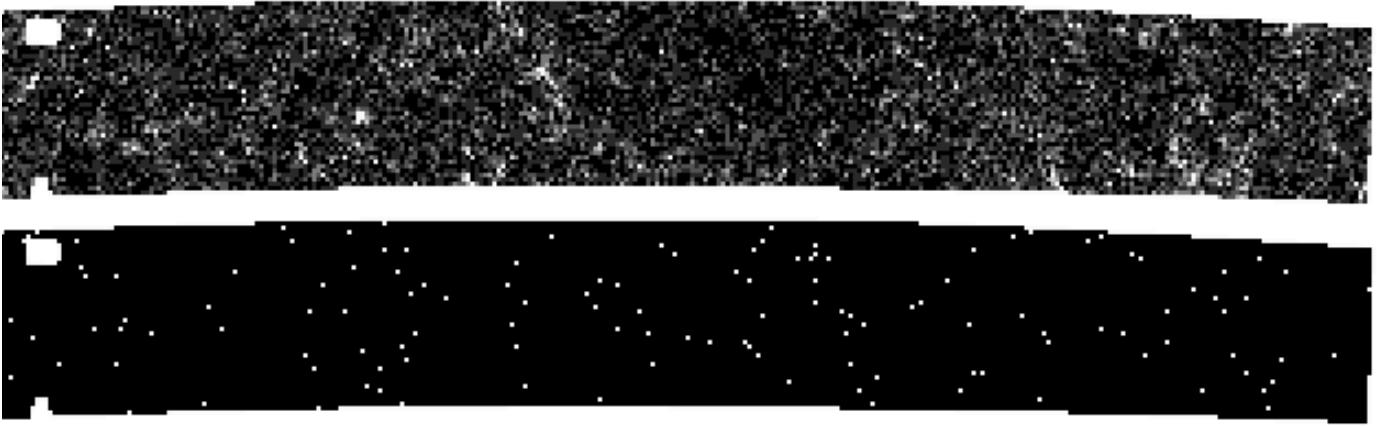}}
}
\caption[Fig2] {\label{pixelMap2} 
Zoom over the central part of the SDSS EDR/N (bottom slices in
previous Figure).}
\end{figure*}

The 1st and 3rd slices in Figure \ref{pixelMap} shows the EDR/S and EDR/N
pixel maps for $i'<18.5$ and  $7$ arc-minute
resolution.  Note the "barrel" shape in the EDR/N. As far as we have
been able to find out, this seems an unreported artifact in the EDR redshift
sample release which does not seem to include redshifts for secondary targets
in this equatorial strip (see Stoughton et al 2002 for details). This is not a
problem in our analysis other than we are missing a good fraction of the
EDR/N because there are no matching redshifts for the QSOs.
The top slice in Figure \ref{pixelMap2} shows a zoom over the central region of
EDR/N.

We recovered the QSO sample form the SDSS/EDR as described in detailed in
Schneider et al (2002, see also Richards \etal 2002).  We use point-spread
function magnitudes for QSO (which will be quoted with a subindex: $i'_Q$)
and Petrosian magnitudes for galaxies (which will be
quoted without any subindex: ie $i'$), as indicated by above references.  
We have considered two QSO samples. The one obtained
directly from the EDR data archive (ie just with $\rm specobj.specClass=3$)
containing 4275 QSOs, which we call EDR/QSO, and the corrected public QSO
sample presented in Schneider et al (2002), which we called SDSS/QSO and
contains 3814 QSOs. Almost all SDSS/QSO is contained in EDR/QSO.  SDSS/QSO
contains a handful of additional BALQSOs and some quasars that were identified by
eye during the checking process.
SDSS/QSO excludes most of the low-redshift AGN ($M_i<-23$)
and a small number of narrow-line QSO included EDR/QSO, but this
makes little different for our sample which has $z>0.8$.

Out of the EDR/QSO sample we impose two further cuts: $\rm specobj.zStatus>2$
to exclude failed or inconsistent redshifts and $\rm specobj.zConf>0.7$ to
exclude redshifts with confidence less than $70\%$ (see Stoughton \etal 2002
for details). In both samples we restrict our analysis to $0.8 < z<2.5$ where
the QSO distribution is compact and homogeneous (the lower cut avoids overlap
with the galaxy sample). After these cuts the main difference between EDR/QSO
and SDSS/QSO are the few missing radio QSO and a few narrow-band emission
QSO's. Over $97\%$ of the QSOs are the same and have the same parameters.
Note that according to Schneider \etal 2002, QSO selection criteria were
changed systematically during the EDR runs resulting in large variation on the
number of QSOs per field. Nevertheless both samples are 
photometrically completed to better than $90\%$ in
$16.5<i'_Q<19.1$\footnote{Recall that $i'$ are 'raw' magnitudes. The
  extinction corrected limit is $i^*<19.0$.}  (Schneider \etal 2002, Stoughton
\etal 2002). We will therefore restrict our analysis to $16.5<i'_Q<19.1$ and
further compare results for the two versions of the target selection that are
relevant to our sky coverage: v2.2a and v2.7 (see Stoughton \etal 2002).

Number counts and the redshift distribution are shown in
Fig.\ref{edrqsoccounts}.  Note the sharp break in the QSO number counts at
$i'_Q \simeq 19$. The short-dashed line shows a power law with $10^{0.4 m}$
(which corresponds to $\alpha =1$ in Eq.[\ref{deltamu1}]).  As lensing
magnification should be negative for shallower slopes, we cut the QSO sample
to $i'_Q < 18.8$ (which contains $952$ QSOs for $0.8< z < 2.5$ within
our mask) and use the fainter cut
$i'_Q > 18.8$ data (which contains $671$ QSOs within our mask) for comparison.  Thus, unless
stated otherwise, QSO are selected with $16.5<i'_Q < 18.8$, which for brevity
will be sometimes refered as $i'_Q < 18.8$.

Right panel in Fig.\ref{edrqsoccounts} shows how the redshift distribution for
$i'_Q < 18.8$ of EDR/QSO (histogram) and SDSS/QSO (dashed
line) are almost identical, with a mean redshift of
$\zbar_{QSO} = 1.67$. The redshift distribution 
 for $i'_Q < 19.1$ is very similar (with just
higher surface density).  Here we will only present results based on EDR/QSO.
We have done all the analysis for SDSS/QSO and find identical results in all
cases.  We choose to present EDR/QSO because it has been automatically
produced (in the same pipeline as the galaxies) and because we are not concerned
with the astrophysical nature of  the QSO  in the sample but rather with
having a well defined photometric sample of distant objects.

Figure \ref{pixelMap} compares the galaxy pixel maps $i'<18.5$ with the QSO
$i'_Q<18.8$ distribution in the same portion of the sky (QSOs are the 2nd and
4th slices below the corresponding EDR/N and EDR/S galaxies).  Figure
\ref{pixelMap2} shows a zoom over the central region of EDR/N.
Note that there is no apparent correlation between QSO and
galaxies. As we will see below, most of the signal we are seeking for is
hidden below the pixel resolution of this map.

\begin{figure*} 
\centering{
{\epsfxsize=18cm \epsfbox{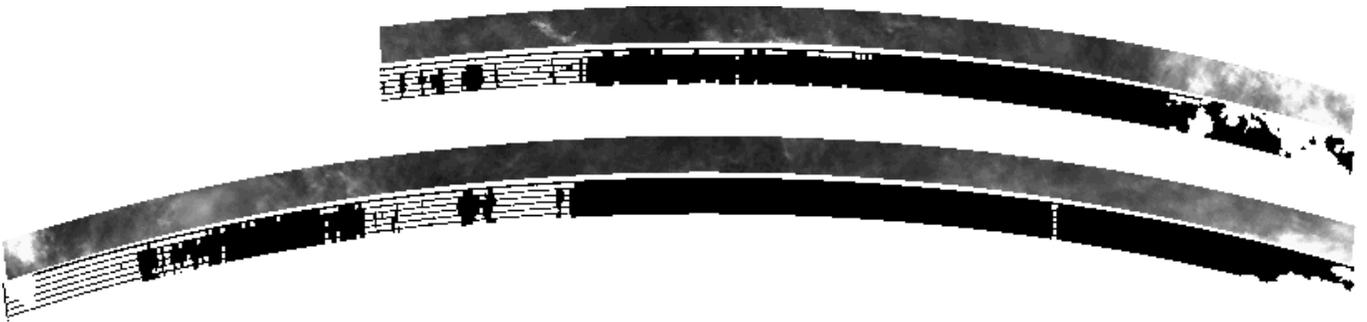}}
}
\caption[Fig2] {\label{Mapext} 
Pixel maps  of 
galactic absorption (Schlegel \etal 1998) over the EDR/N (top two slices)
and EDR/S region. We also show below each extinction map the EDR/SDSS mask
of pixels with less than $0.2$ mag extinction in $i'$ and 
less than $2"$ seeing.}
\end{figure*}

\begin{figure*}
\centerline{\epsfysize=8truecm
\epsfbox{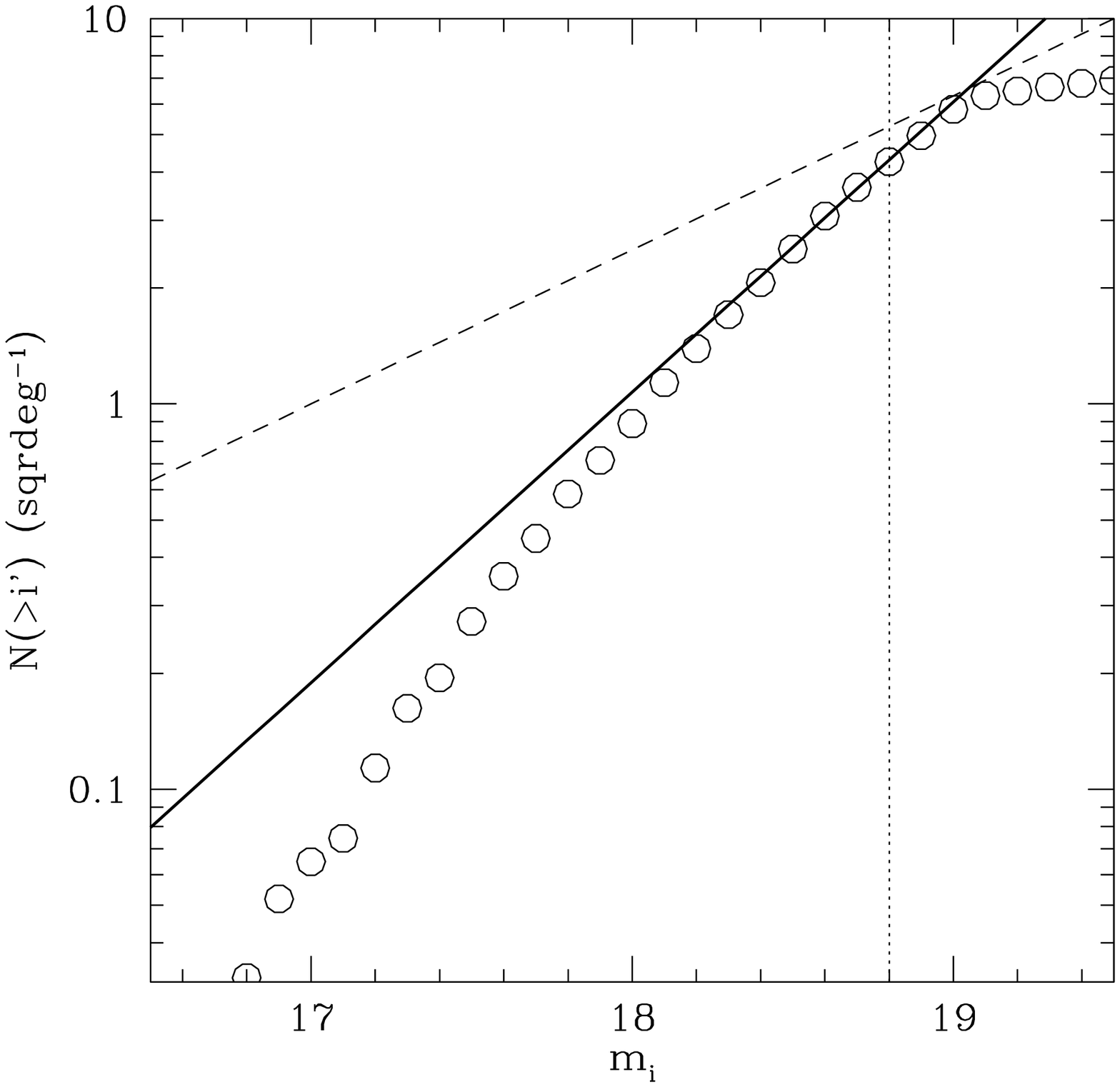}\epsfysize=8truecm
\epsfbox{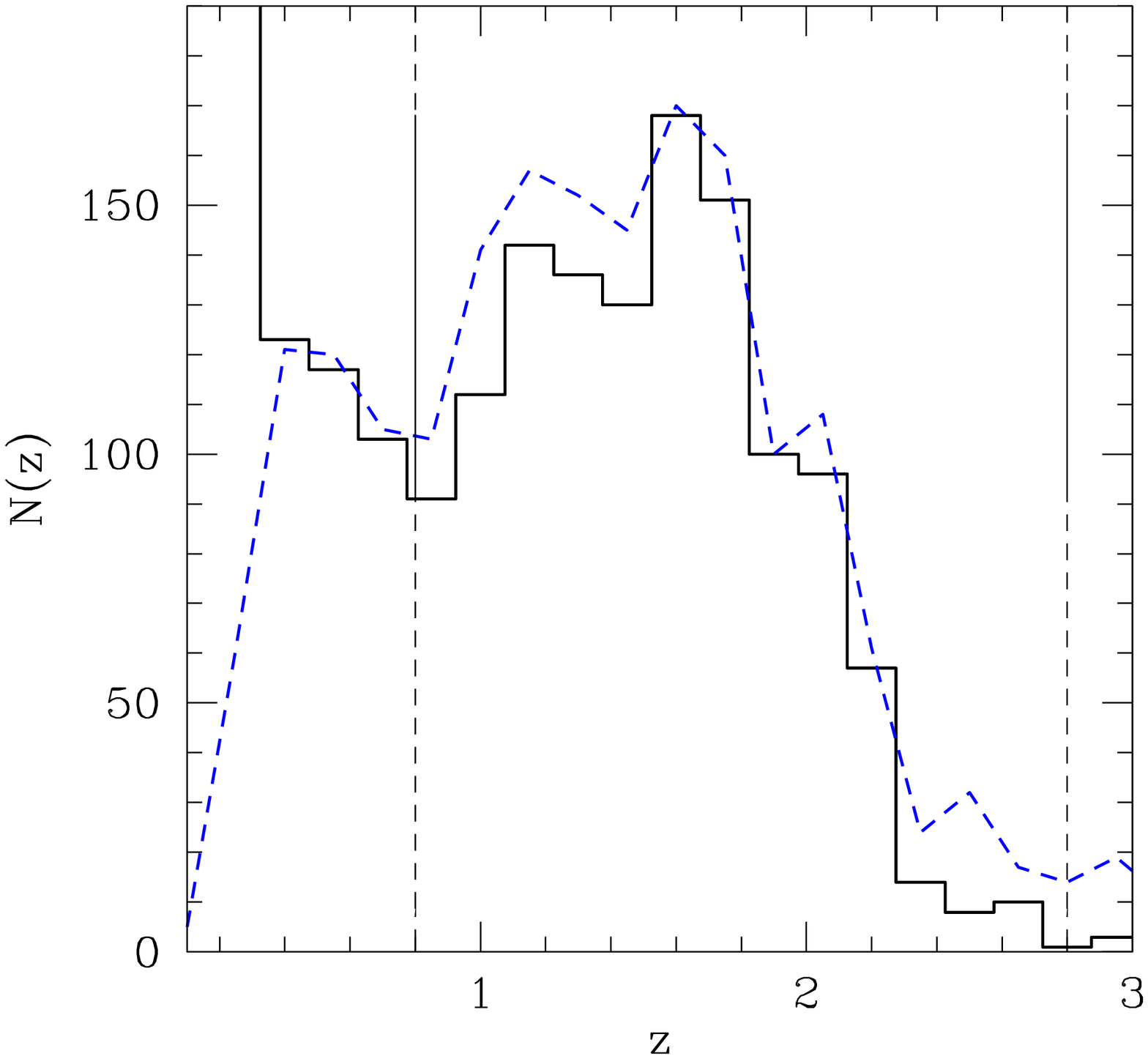}}
\caption[Fig2]{\label{edrqsoccounts}
{\sc Left:} Integreted QSO number counts per unit magnitude and area.
{\sc Right:} Redshift distributions for the EDR/QSO sample (continuous line)
and the SDSS/QSO sample (dashed lines). In both case $16.5<i_Q'<18.8$.}
\end{figure*}

\section{Correcting for Galactic extinction}

Absorption by Galactic dust can lead to a positive correlation between
galaxies and high-z QSOs (see Norman \& Williams 1999).  Schlegel \etal (1998)
extinction maps have a significant differential extinction $E(B-V) \simeq
0.02-0.03$ even at the poles. Thus, the extinction correction has a large
impact in the number counts for a fix magnitude range.  The change can be
roughly accounted for by shifting the mean magnitude ranges by the mean
extinction.  Despite this, extinction has little impact on clustering, at
least for $r'<21$ (see Scranton \etal 2001 and also Tegmark et al. 1998).
This is fortunate because of the uncertainties involved in making the
extinction maps and its calibration.  Moreover, the Schlegel \etal (1998)
extinction map only has a 6'.1 FWHM, which is much larger than the individual
galaxies and QSO we are interested on. Many dusty regions have filamentary
structure (with a fractal pattern) and large fluctuations in extinction from
point to point (ie see Fig.\ref{Mapext}). 
One would expect similar fluctuations on smaller (galaxy size)
scales, which introduces further uncertainties to individual corrections.  As
we are looking for a very low signal (of order of $1\%$) we have to be very
careful with small systematic effects, such as extinction.

The net effect of extinction is to produce 
a magnitude absorption $A$ which translates into
a change in the local galaxy surface number density $N$:
\beq
\Delta_A \equiv {\Delta N \over{N}} \simeq \alpha A
\eeq
where $\alpha$ is the slope of the number density counts: $N \simeq 10^{\alpha m}$
as a function of magnitude $m$. If the absorption $A$ were a constant, this
 will would only change the mean number density in an uniform way. Unfortunately 
extinction is highly variable and produces density fluctuations in the sky:
\beq
\delta_A \equiv \Delta_A - <\Delta_A> = \alpha \left(A- <\Delta_A> \right)
\eeq
We thus see that extinction will introduce spurious number density fluctuations
 in both the galaxy and the QSO distributions:
\bea
\deltab_G &=& \delta_G + s_G ~\delta_A \\
\deltab_Q &=& \delta_Q + s_Q ~\delta_A 
\eea
where $\deltab_G$ and  $\deltab_Q$
stand for the total observed fluctuations (as opposed to the
intrinsic ones, $\delta_G$ and $\delta_Q$) 
and $s_G$ and $s_Q$ are constant numbers
 characteristic of each population. From the above analysis we can see that
extinction will introduce artificial cross-correlations between the galaxy and
QSO populations, even when they are intrinsically uncorrelated. 
\footnote{A similar argument and analysis can be made for other systematics, such as
seeing variations.}
One can correct for this type of effects by using the absorption maps.
We can  calculate the cross-correlations:
\bea
< \deltab_G~\delta_A> &=&  s_G ~<\delta^2_A > \\
< \deltab_Q~\delta_A> &=&  s_Q ~<\delta^2_A > \\
< \deltab_G~\deltab_Q> &=& < \delta_G~\delta_Q> + s_G ~s_Q ~<\delta^2_A >
\eea
where we have assumed that the intrinsic galaxy and QSO positions are uncorrelated
with extinction. We thus have:
\beq
< \deltab_G~\deltab_Q> = < \delta_G~\delta_Q> + {< \deltab_G~\delta_A> < \deltab_A~\deltab_Q>\over{<\delta^2_A >}}
\label{wGQext}
\eeq
where $ < \delta_G~\delta_Q>$ is the intrinsic cross-correlation
(eg from cosmic magnification).
As we can measure all $< \deltab_G~\delta_A>$, $< \deltab_Q~\delta_A>$ and
$<\delta^2_A >$ from the maps, the above expression allow us to correct the
galaxy-QSO cross-correlation for extinction.  

To test this model we will
consider 3 types of magnitude corections: {\it raw magnitudes} (ie uncorrected from
extinction) : $i'$; extinction $A$ corrected magnitudes: $i^*=i'-A$; and extinction 
over-corrected magnitudes: $i'+A$. The later case will give us a good diagnostic
to test our method to estimate the contamination in the galaxy-QSO cross-correlation.

\begin{figure*} 
\centerline{\epsfysize=8truecm
\epsfbox{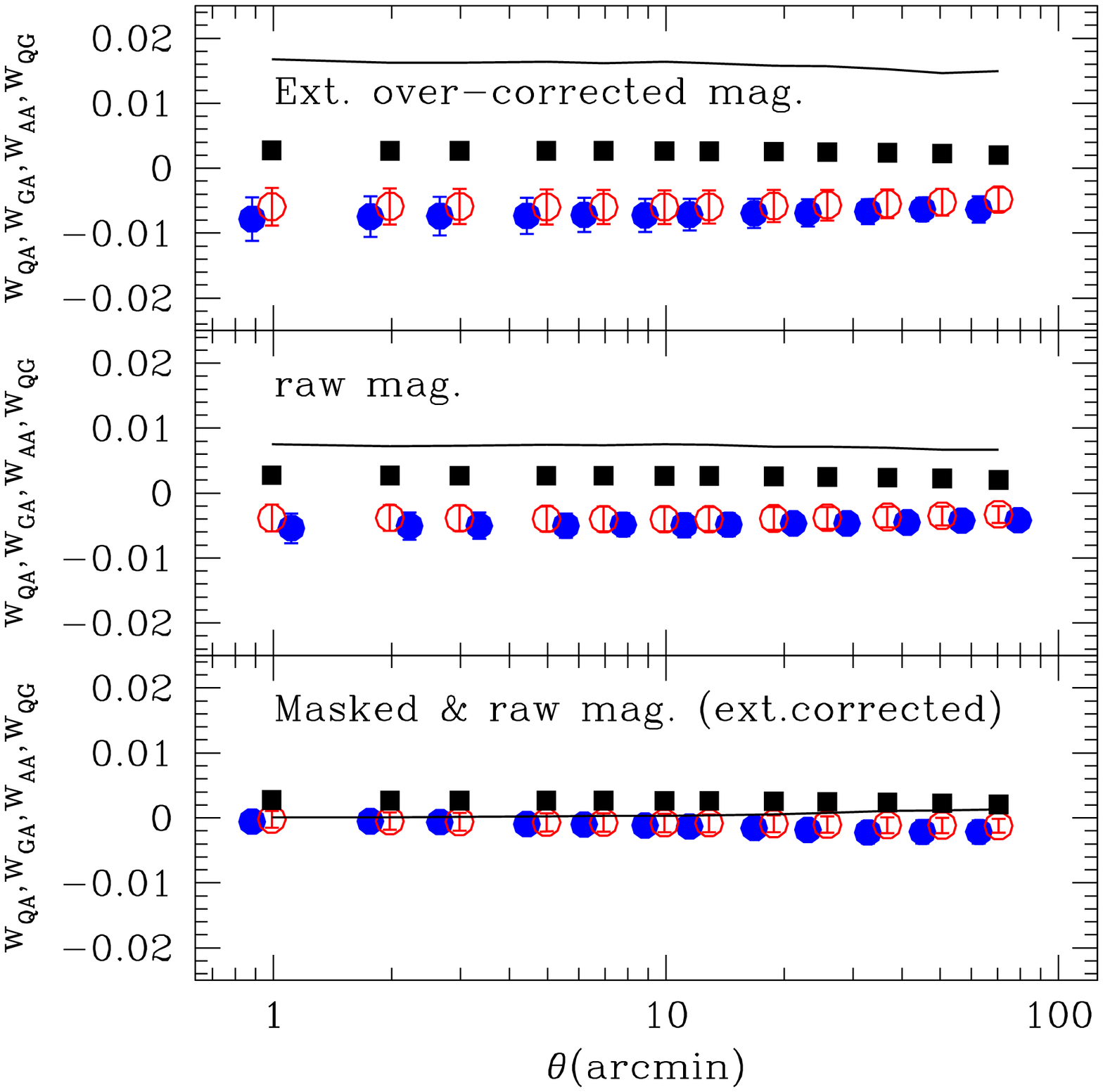}\epsfysize=8cm 
\epsfbox{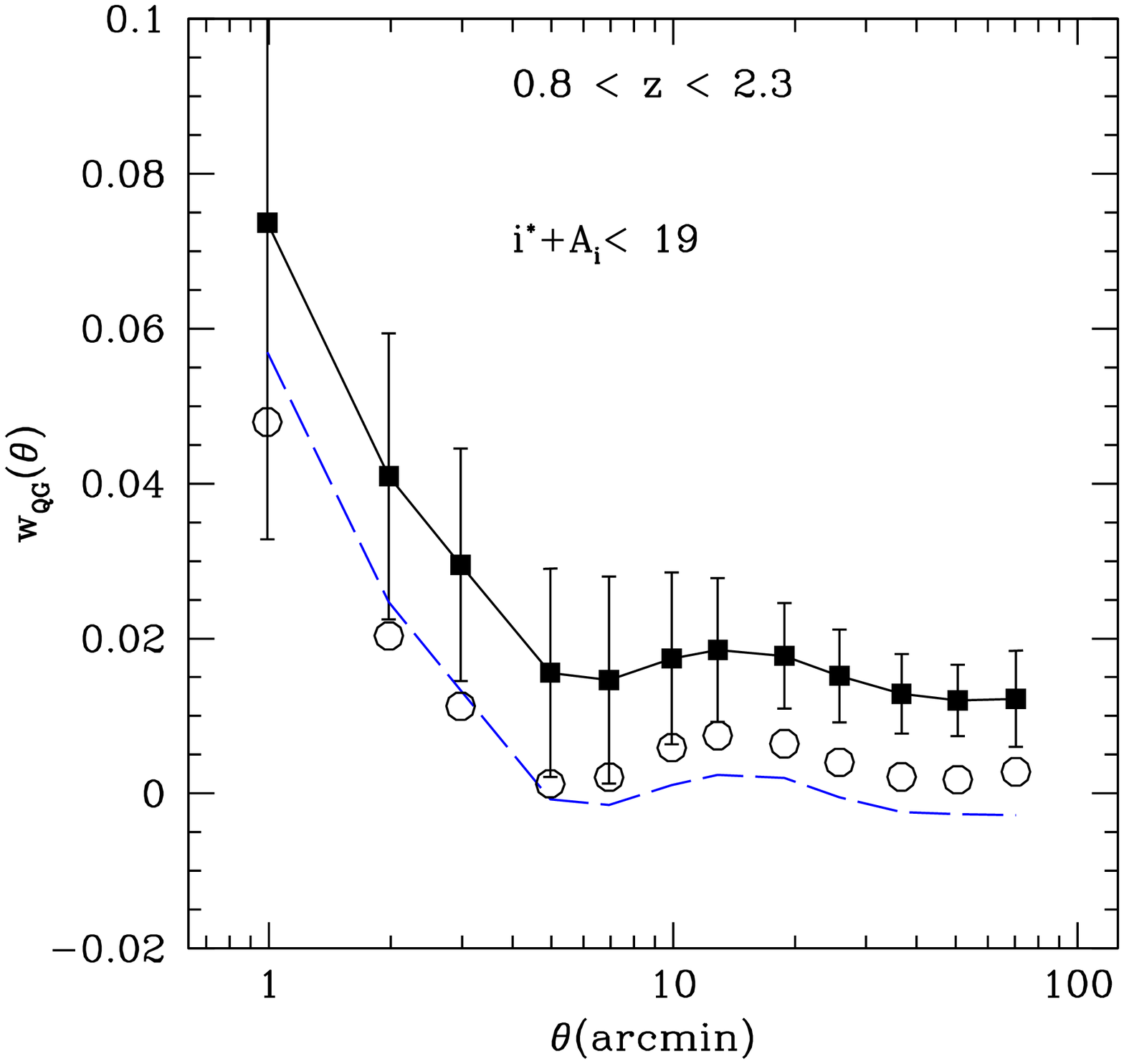}}
\caption[Fig2]{\label{w2z}
{\sc Left Panel:} The density variance in the extinction
$<\delta^2_A >$ (closed squares), compared with galaxy-extinction
 cross-correlation  $< \deltab_G~\delta_A>$ (open circles)
and the QSO-extinction  cross-correlation 
$< \deltab_Q~\delta_A>$ (closed circles).
The continuous line shows the contamination
in the galaxy-QSO cross-correlation as predicted by Eq.[\ref{wGQext}]. 
The top, middle and bottom panels correspond to extinction over-corrected
magnitudes, raw magnitudes and results for raw magnitudes
with extinction and seeing mask (which agrees well with the extinction
corrected result). 
{\sc Right Panel:} galaxy-QSO cross-correlation  $w_{GQ}$ as a function of
cell radius $\theta$, for objects selected
with extinction over-corrected magnitudes $i^*<19$ (closed squares and continuous
line) and raw magnitudes $i'<19$ (open circles). The dashed line shows the
prediction in Eq.[\ref{wGQext}] using $<\delta^2_A >$,
$< \deltab_G~\delta_A>$ and $< \deltab_Q~\delta_A>$  cross-correlations
shown in the top left panel.}
\end{figure*}

The results for these quantities are shown in the left panel of Fig.\ref{w2z}.
Note how $<\delta^2_A >$ (shown as closed squares) is quite flat. This is because
of the lack of resolution on scales $\theta < 10'$. On larger scales, Schlegel
\etal (1998) find that the angular spectrum of their absorption maps fit well
$P(k) \propto k^{-2.5}$ at all scale which corresponds to $<\delta^2_A>
\propto \theta^{+0.5}$, so that fluctuations grow only slowly with scale.

Fig.\ref{w2z} shows the prediction in Eq.[\ref{wGQext}] as a continuous line,
which is close to $2\%$ for the top case (in all cases we are using $i<19$
pixel maps for both QSO and galaxies).  As can be seen in the figure, when we
over-correct the magnitudes for extinction (top panel) the correction is
larger than when we use raw (uncorrected) magnitudes (shown in the middle
panel), as expected.  This illustrates how extinction could induce
cross-correlations in the QSO and galaxy maps.  When we apply the right
(negative) extinction correction to the magnitudes (bottom left panel) we find
a negligible correlation within the error-bars. This result agrees well with
that of using the raw magnitudes with and extinction mask (ie masking out the
high extinction pixels, see below). Either of this two last options is
acceptable for our analysis as the residual cross-correlation is negligible
given the error-bars.

In the right panel of Fig.\ref{w2z}
we show the galaxy-QSO cross-correlation $w_{GQ}$ for maps
with extinction over-corrected magnitudes $i^*<19$ (closed squares)
with  $w_{GQ}$ for maps with  raw $i'<19$ (open squares). The dashed
line shows again $w_{GQ}$ with extinction over-corrected magnitudes 
after  subtracting the correction in   Eq.[\ref{wGQext}] (ie continuous
line in the top left panel of Fig.\ref{w2z}). The agreement is quite good
indicating that there is only marginal contamination of extinction in
$w_{GQ}$ when we use raw magnitudes (in agreement with the middle left panel
of  Fig.\ref{w2z}). 

These results 
gives us some confident that we can control this type of contamination
in  $w_{GQ}$ (and also provides a further test to our cross-correlation codes).

\subsection{Extinction and seeing mask}

Following Scranton et al (2002), pixels (in 6' resolution) with larger mean
seeing or larger mean extinction than some threshold value are masked out from
our analysis.  The final mask is the product of this seeing and extinction
mask (ie shown in Fig.\ref{Mapext}) by the sample boundary (shown in
Fig.\ref{pixelMap2}).  We have tried different thresholds following the
analysis of Scranton et al (2001).  As a compromise between precision and area
covered, unless stated otherwise, we use 0.2 maximum extinction and seeing
better than 1.8' for $r' <22$ and $i'<22$. For $r'<19$ we relax the
seeing cut to 2 arc-sec to include more galaxies. For these brighter galaxies
2 arc-sec provides very good photometry for clustering analysis (see Scranton
\etal 2001).

Bottom left panel in  Fig.\ref{w2z} shows the cross-correlation 
results after applying the extinction mask over raw $i'<19$ magnitudes.
The resulting contamination in $w_{GQ}$ is negligible.

\section{Cross-correlation measurements}

For our statistical analysis, we will use moments of counts in cells, eg
the variance:
\beq
w_{GG}(\theta) = <\delta^2_G(\theta) >
\eeq
where $\delta_G \equiv n_G/<n> -1$ are number density fluctuations 
on cells of size $\theta$ 
(larger than the pixel map resolution)
and $<n>$ is the mean number of galaxies in the cell. The average
$<...>$ is over angular positions in the sky.  We follow
closely Gazta\~naga (1994, see also Szapudi \etal 1995) and use the same
software and estimators here for the SDSS. This software  have been
tested in different ways and the results confirmed by independent studies (eg
see Szapudi \& Gazta\~naga 1998).

For the cross-correlation we use (see also MBM02):
\beq
w_{GQ}(\theta) = <\delta_G(\theta) \delta_Q(\theta) >
\eeq
Note that this is different from the 2-point cross-correlation:
$w_2(\theta_{12}) = <\delta_G(\vtheta_1) \delta_Q(\vtheta_2) >$,
where $\theta_{12}=|\vtheta_2-\vtheta_1|$. In our case,
both cells are at the same location in the sky, 
and the scale dependence comes from changing the cell size.
In fact, $w_{GQ}$ is just an area average over $w_2$, and its
amplitude at scale $\theta$ is typically $20-30 \%$ 
higher that $w_2$ at $\theta_{12} \simeq \theta/\sqrt{\pi}$
(see Fig.1 in Gazta\~naga 1994). Also note that shot-noise
cancels out for cross-correlation  $w_{GQ}$, but not for
 $w_{GG}$  and  $w_{QQ}$ where the variance needs to be
shot-noise corrected (eg Gazta\~naga 1994).

The reason for using the variance $w_{GQ}$ rather than the 2-point
$w_2(\theta_{12})$ function is twofold. Firstly one expects the variance to
provide better signal-to-noise ratios when the correlation signal is small,
just because $w_{GQ}$ is an integrated quantity.  Secondly, M\'enard \etal
(2002) have argued that $w_{GQ}$ provides a better estimator to improve the
accuracy of the comparison with the theoretical predictions in the weak
lensing regime. We have in fact performed a calculation of the
$w_2(\theta_{12})$ cross-correlation over the same data, using the simple
counts-in-cell estimator for $w_2(\theta_{12})$ (eg Eq.[36] in Gazta\~naga
1994). The results were dominated by noise (sampling errors) and we have
therefore decided to just concentrate our efforts on $w_{GQ}$.

We will also measure the galaxy-galaxy-QSO 3rd order moment:
\beq
w_{GGQ}(\theta) = <\delta_G(\theta) \delta_G(\theta) \delta_Q(\theta)>
\eeq

and the galaxy-galaxy variance around QSOs:
\beq
w_{GG;Q}(\theta) = <\delta_G(\theta) \delta_G(\theta) >_{QSO}
\eeq
The extra or excess variance is defined as (MBM02):
\beq
\Delta(\theta) \equiv w_{GG;Q}(\theta) - w_{GG}(\theta)
\eeq
which turns out to be a measure of the 3-point function
$w_{GGQ}(\theta) \sim \Delta(\theta)$ (Fry \& Peebles 1980, MBM02).

Errors are obtained from a variation of the jackknife error scheme proposed by
Scranton et al (2001, Eq.[10]). The sample is divided into $N$ separate
regions on the sky, each of equal area (in our case they are cuts in RA
such that the number of pixel cells is equal in each region). The analysis
is performed $N$ times, each leaving a different region out. These are called
(jackknife) subsamples, which we label $L=1 \dots N$. The estimated
statistical covariance for $w_{GQ}$ for scales $\theta_i$ and $\theta_j$
is then:
\bea
Covar(\theta_i,\theta_j) &\equiv&
<\Delta w_{GQ}(\theta_i)~\Delta w_{GQ}(\theta_j) > \label{jack}
\\ \nonumber
&=& {N-1\over{N}} \sum_{L=1}^{N} \Delta w_{GQ}^L(\theta_i) \Delta
w_{GQ}^L(\theta_j) \\
\Delta w_{GQ}^L(\theta_i) &\equiv& w_{GQ}^L(\theta_i) - \widehat{w_{GQ}}(\theta_i)
\eea
where $w_{GQ}^L(\theta_i)$ is the cross-correlation measure in the 
$L$-th subsample ($L=1
\dots N$)  and $\widehat{w_{GQ}}(\theta_i)$ is the mean value for the $N$ subsamples.
The case $i=j$ gives the error variance. Note how if we increase the number of
regions $N$, then jackknife subsamples are larger and each term in the sum is smaller. 
We have checked that the resulting covariance gives a stable answer for
different $N$. 

These errors has been shown to be 
reliable when tested against simulations in Zehavi et al (2002) and
have the great advantage of being model independent.  In our implementation we
use $N=20$  (but we have also tried $N=10$ and $N=40$)
 independent regions of equal size.

\subsection{Variance in the galaxy-QSO correlation}

\begin{figure*} 
\centerline{\epsfysize=8truecm
\epsfbox{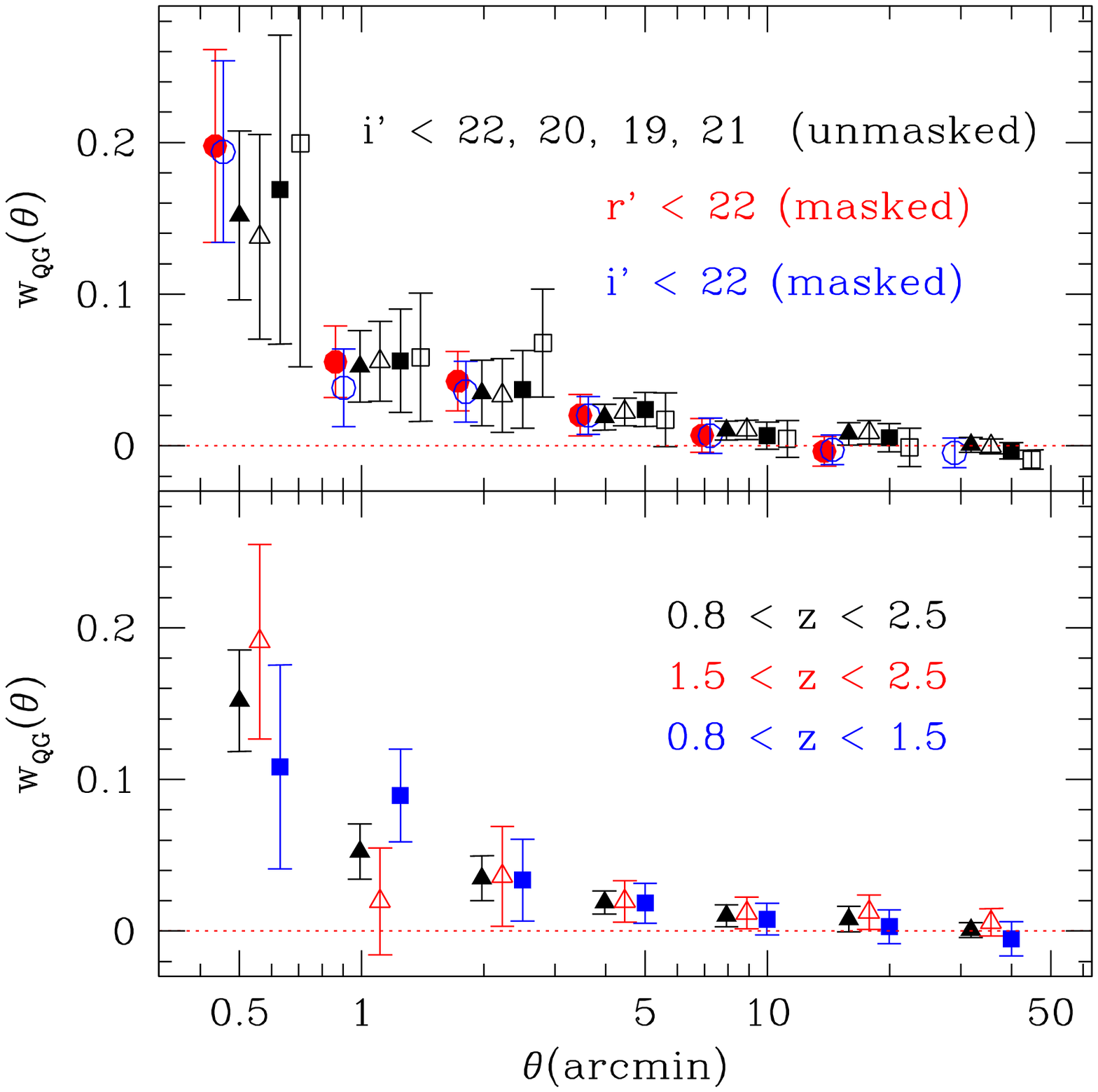}\epsfysize=8truecm
\epsfbox{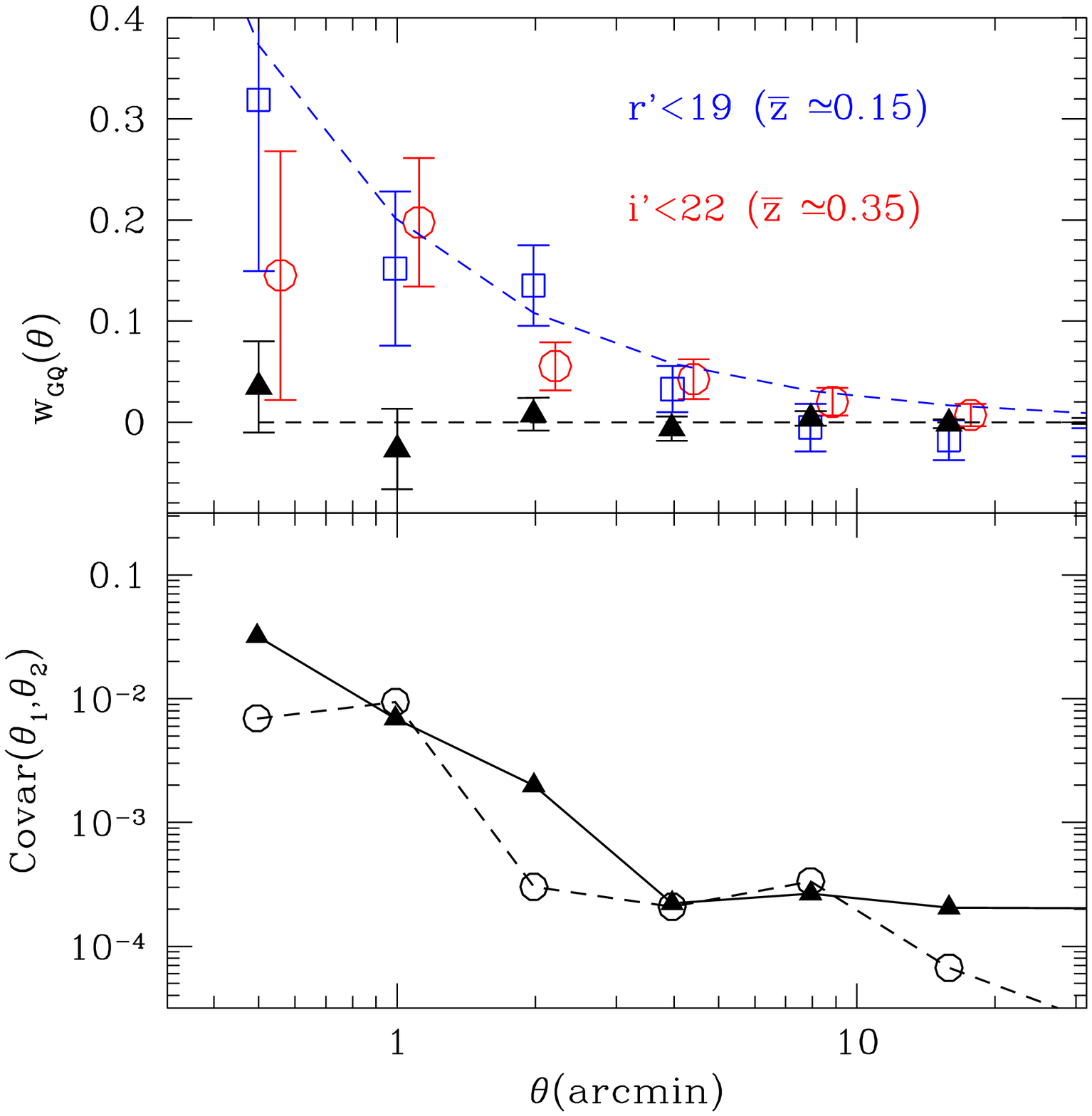}}
\caption[Fig2]{\label{w2mz}
{\sc Top Left panel:} 
Galaxy-QSO cross-correlation $w_{GQ}$ (symbols with error-bars)
for different galaxy samples as a function of cell radius
(which is slightly shifted in each case to avoid overlaps). 
Closed and opened circles correspond to
$i'<22$ and $r'<22$ with pixels masked (less than $0.2$ mag extinction and 
more than $2"$ seeing). Closed triangles, open triangles,  closed squares and open
squares, show the unmasked results for galaxies with limiting fluxes
of $i'=22, 21, 20, 19$ respectively. 
{\sc Bottom Left panel:}  $w_{GQ}$ for $r'=22$ galaxy sample and different
redshift bins in the QSO sample: $0.8 <z <1.5$ (closed squares) and
$1.5 <z <2.5$ (open triangles) and $0.8 <z <2.5$ (closed triangles).
{\sc Top Right panel:} Circles and squares  with error-bars show
galaxy-QSO cross-correlation $w_{GQ}$ 
for $r'<19$ ($\zbar \simeq 0.15$)
and for $i'<22$ ($\zbar \simeq 0.35$). 
Closed triangles show  $w_{GQ}$ in  $i'<22$ when we randomize the galaxy
counts. For comparison we also show $w_{GQ} = 0.2 \theta^{-0.9}$ as a dashed line.
{\sc Bottom panel:} Covariance $Covar(\theta_1,\theta_2)$ for
$\theta_1=0.5'$ (continuous line) and $\theta_1=1.0'$ 
(dashed line) as a function of $\theta_2$.}
\end{figure*}

The top-left panel of Figure \ref{w2mz} compares the measured values of
$w_{GQ}(\theta)$ for the different magnitude bins cuts $r'=19, 20,21, 22$
with unmasked pixels with the results for
$r'<22$ and $i'<22$ with the extinction and seeing mask. 
All cases are roughly consistent with each other.
Differences will be studied in section \S 5, the point here is that there is a
significant signal in all bands (very similar results are find for cuts in
$g'$). Bottom left panel compares
$w_{GQ}(\theta)$ when we separate the QSO in high and low redshifts.

The top-right panel of Figure \ref{w2mz}
shows in more detail the results for $r'<19$ ($\zbar
\simeq 0.15$, squares) and for $i'<22$ ($\zbar \simeq 0.35$, circles). 
The later have been
shifted in angular scale (up by $\simeq 2.2$) to match the $r'<19$ depth.
Closed triangles show the $i'<22$ results after
randomizing the angular position of the galaxy counts-in-cells. This produces
a distribution with identical variance, skewness and higher order moments,
but which should not be correlated to the QSO sample. Thus, 
we  expect  $w_{GQ}=0$ in this case, exactly as found within the error-bars.
This provides a test for our code and statistics against systematics such as
boundary problems in the comparison. Also provides an indication that the
errors are not underestimated.

The bottom-right panel in  Figure \ref{w2mz} shows  
the covariance matrix Eq.[\ref{jack}]
 for $\theta_1=0.5$ (continuous line) and $\theta_1=1.0$ (dashed line) 
from the jackknife estimator with $N=40$ (values for $N=10$ are comparable but
the scatter is larger). As shown in the Figure the dominant contribution comes
 from the diagonal terms. This is because we are using
 well separated bins (each cell is 4 times larger than the previous one).
 We neglect the off-diagonal errors in a first
 interpretation of the data, but we note that a stronger covariance arises
from bins $\theta_2 <\theta_1$ than for $\theta_2 >\theta_1$, as shown in
the Figure for the $\theta_1=1.0'$ case.

All cases on the right panel of  Figure \ref{w2mz} correspond to maps 
with extinction 0.2 and 2" seeing mask for  $r'<19$ and 1.8" seeing mask
for  $i'<22$ (and $i'_Q < 18.8$).

Assuming no covariance, the significance of the detection (against zero
correlation) for the first 5 points in top-right panel of Figure \ref{w2mz} is
about 4-sigma in $r'<19$ and 8-sigma in $i'<22$.

\begin{figure*} 
\centerline{\epsfysize=8truecm
\epsfbox{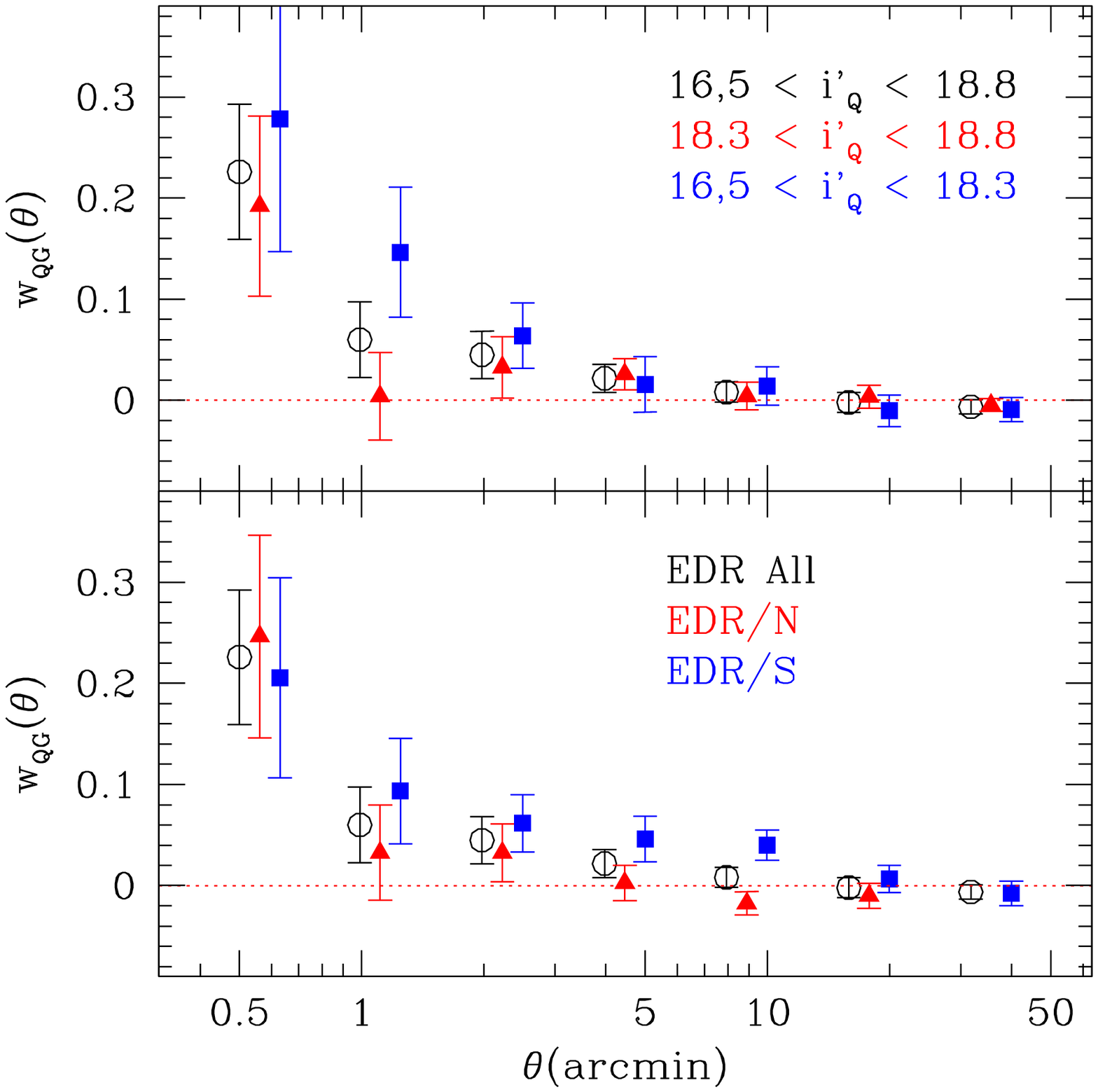}\epsfysize=8truecm
\epsfbox{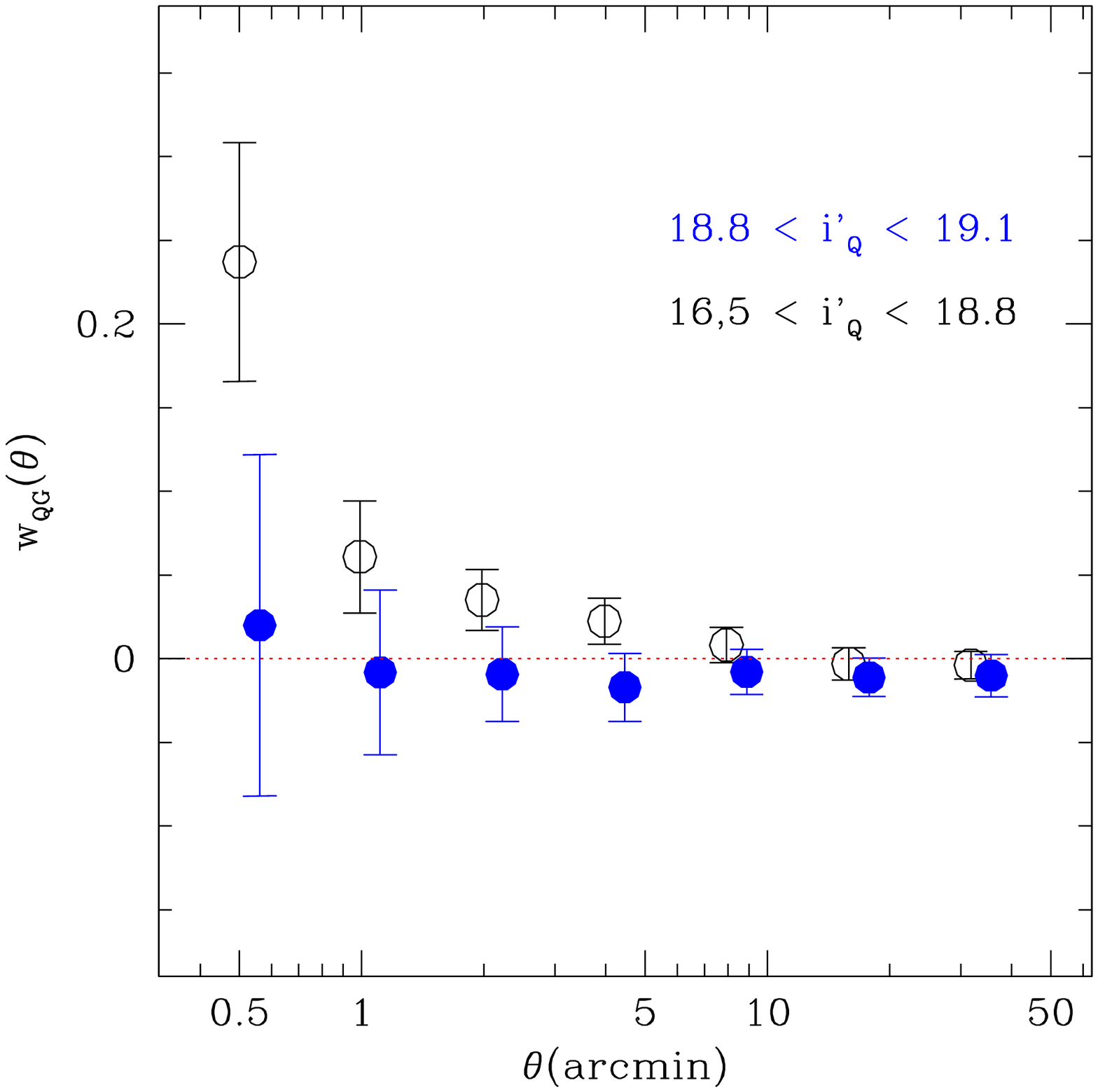}}
\caption[Fig2]{\label{w2icut}
{\sc Top Left panel:} 
galaxy-QSO cross-correlation $w_{GQ}$
for $i'<22$ galaxies  with
different QSO sub-samples: 
$16.5 <i'_Q<18.3$ (closed squares),
$18.3 <i'_Q<18.8$ (closed triangles)
and the combined $16.5 <i'_Q<18.8$ (open circles).
{\sc Bottom Left panel:} Comparison of the mean  $w_{GQ}$ (open circles)
with the values in the North (closed squares) and South 
(closed triangles) EDR subsamples (in all cases
$i'<22$ galaxies and  $i'_Q <18.8$).
{\sc Right panel:} Galaxy-QSO
cross-correlation $w_{GQ}$ (symbols with error-bars)
for $i'<22$ galaxies  with
different QSO sub-samples: $16.5 <i'_Q<18.8$ (open circles) 
and $18.8 <i'_Q<19.1$ (closed circles).}
\end{figure*}

\subsection{Faint and Bright QSOs}

Bottom left panel in Figure \ref{w2icut} shows a comparison of the results in
the EDR/N and EDR/S slices with the combined sample.  The agreement is good
within the errors, with slightly stronger signal in EDR/S.  This provides a
test for different versions in the QSO selection. Runs 752/756 (which
correspond to EDR/N) and 94/125 (which correspond to EDR/S) have different
target selection version: v2.2a and v2.7 respectively.

Top left panel  in Figure \ref{w2icut} shows 
how the masked $i'<22$ galaxies cross-correlated with
QSO subsamples cut at different $i'_Q<18.8$ bands. All results agree
well within the errors with stronger signal for the brighter QSO, as
expected from the steeper slope of the brighter QSOs (see
Fig.\ref{edrqsoccounts}).

Right panel in 
Figure \ref{w2icut} compares $w_{GQ}$
(for masked $i'<22$ galaxies) for our  nominal  $16.5<i'_Q < 18.8$
with the faintest QSOs in EDR/QSO: $18.8<i'_Q < 19.1$.
Note how the fainter QSO sample (closed circles)
show no significant cross-correlation.   This result is in fact 
expected (see \S5 below) if the cross-correlation is truly due to weak-lensing
as $\alpha \simeq 1$ for this faint sample, see Fig.\ref{edrqsoccounts}.

\subsection{Comparison to galaxy-galaxy variance}

\begin{figure*} 
\centerline{\epsfysize=11truecm
\epsfbox{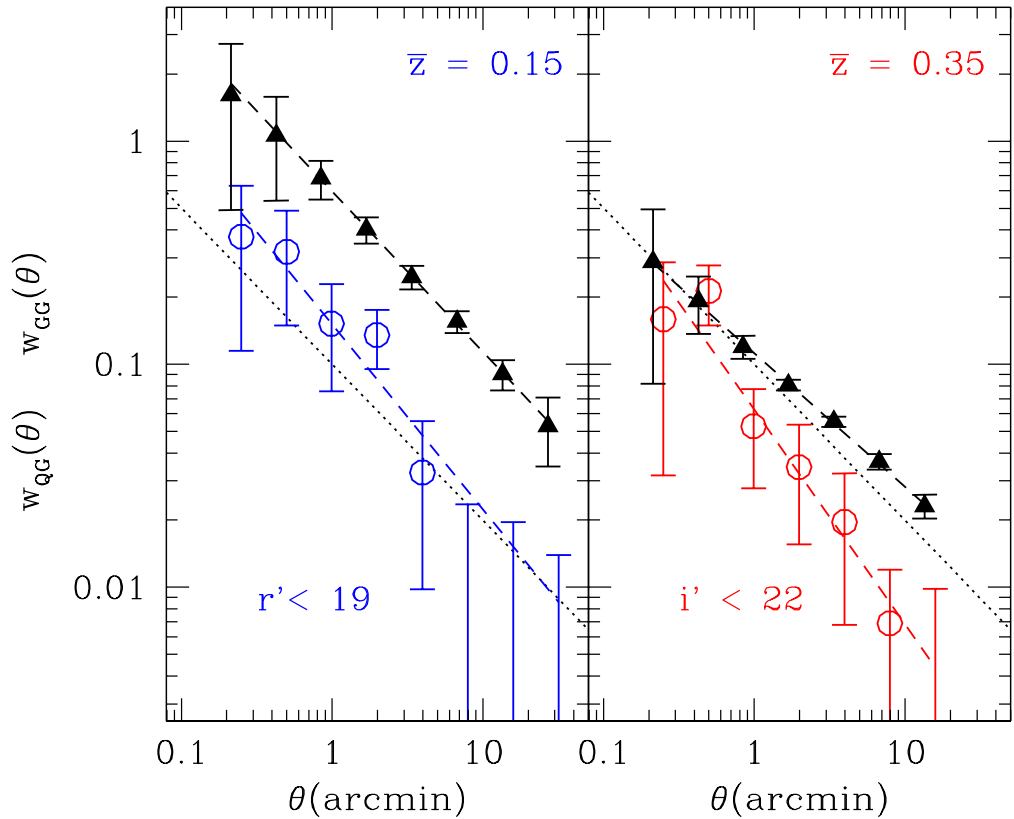}}
\caption[Fig2]{\label{w2gqfit} 
Galaxy-galaxy (closed triangles) and galaxy-QSO (open circles) 
correlation in for bright (left panel) and faint (right panel) galaxies.
Dotted-line shows $w_{GQ} = 0.1 
\theta^{-0.7}$ for comparison, dashed-lines are power-law fits to the data.
Galaxy-galaxy correlations have been shifted slightly to the left  of the
corresponding galaxy-QSO values to avoid
intersection of the error-bars.
}
\end{figure*}

\begin{figure*} 
\centerline{\epsfysize=10truecm
\epsfbox{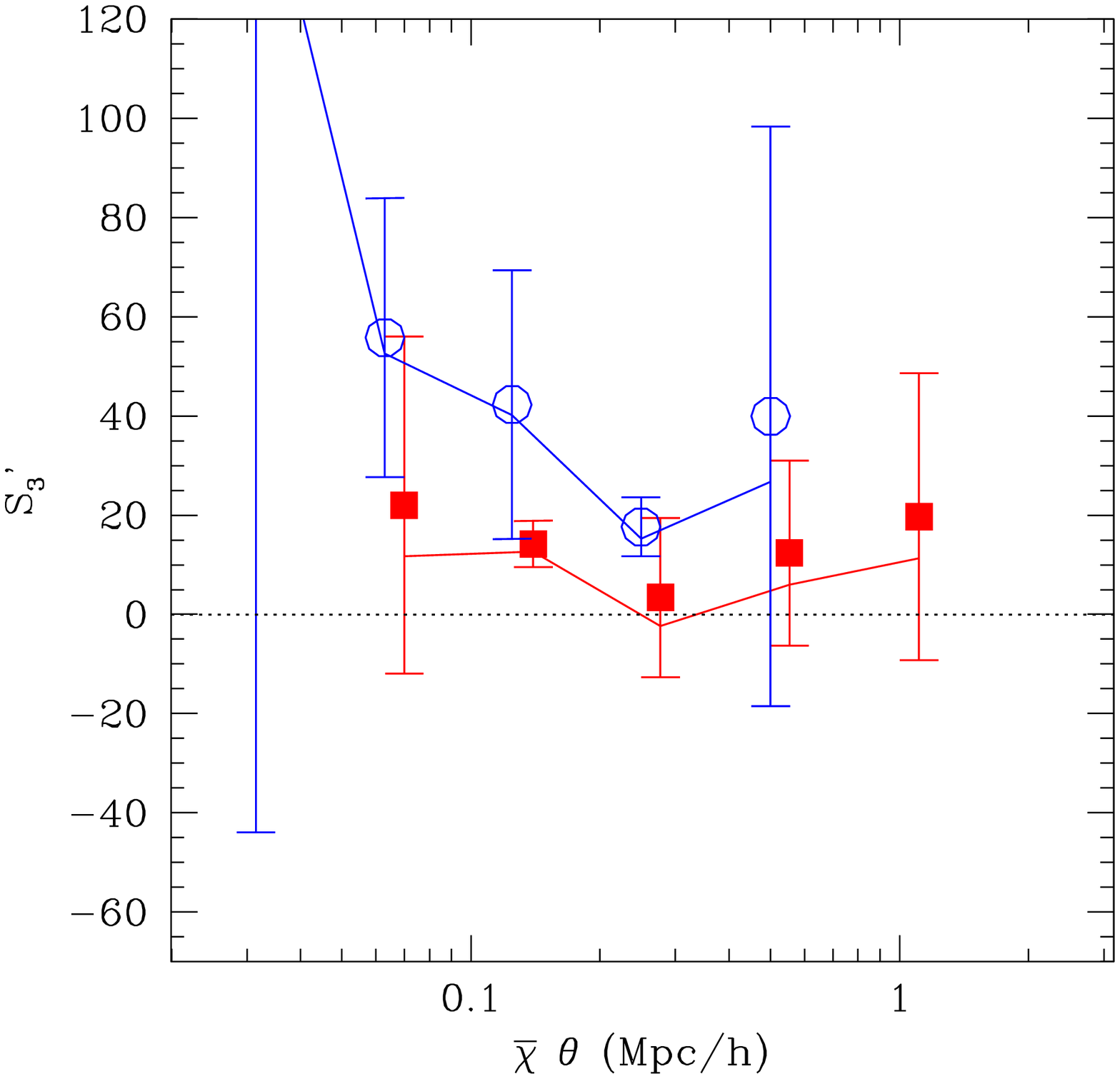}}
\caption[Fig2]{\label{s3edr}
 Pseudo-Skewness Eq.[\ref{s3'}] 
as a function of $\theta$. Square (circles) correspond to
$\zbar \simeq 0.35$ ($\zbar \simeq 0.15$). Continuous lines in each
case correspond to the excess skewness Eq.[\ref{s3''}].} 
\end{figure*}

Fig.\ref{w2gqfit} summarizes  the main observational results in this paper.
We compare $w_{GQ}(\theta)$ (open circles) and $w_{GG}(\theta)$ (closed triangles)
for bright $i'_Q<18.8$ QSOs  with  faint ($r'<19$) and
bright $i'<22$ galaxies samples. The dotted line shows (for comparison)
the power-law $w_{GQ} = 0.1  \theta^{-0.7}$. The  $i'<22$ case is also shown
as closed circles in the top-left panel of Fig.\ref{w2mz}.
 
All data is well fitted to power-laws taking into account the errors (and
neglecting the covariance). We find:

\beq
\begin{array}{ll}
 \beta_{GG} \equiv {d\log{w_{GG}}\over{d\log(\theta)}} =
 & \left\{
\begin{array}{ll}
 -0.72 \pm  0.02 ~~~~ \zbar \simeq 0.15
 \\
 -0.60 \pm  0.03 ~~~~ \zbar \simeq 0.35 
\end{array}
\right.
\end{array}
\label{betaGG}
\eeq

\beq
\begin{array}{ll}
 \beta_{GQ} \equiv {d\log{w_{GQ}}\over{d\log(\theta)}} =
 & \left\{
\begin{array}{ll}
 -0.83 \pm  0.17 ~~~~ \zbar \simeq 0.15
 \\
 -0.96 \pm  0.13 ~~~~ \zbar \simeq 0.35 
\end{array}
\right.
\end{array}
\label{betaGQ}
\eeq

\subsection{The Skewness}

Fig.\ref{s3edr} shows the pseudo-skewness (symbols with error-bars):
\beq
S'_3(\theta) \equiv {w_{GGQ}(\theta)\over{w_{GQ}^2(\theta)}}
\label{s3'}
\eeq
and the  normalized excess skewness (continuous lines):
\beq
S^\Delta_3(\theta) \equiv 
{\Delta(\theta)\over{w_{GQ}^2(\theta)}} =
{w_{GG;Q}(\theta)-w_{GG}(\theta)\over{w_{GQ}^2(\theta)}}
\label{s3''}
\eeq
where $w_{GG;Q}$ is the galaxy variance around QSOs.
It is expected that both quantities, $S'_3$ and $S^\Delta_3$,
should roughly agree on large scales. In fact, when shot-noise
can be neglected  $S^\Delta_3 = S'_3 -1$ (see MBM02). Within the
errors this relation is in good agreement with Fig.\ref{s3edr}.

A fit  of a constant skewness in Fig.\ref{s3edr} gives:

\beq
\begin{array}{ll}
 S_3'  ~=
 & \left\{
\begin{array}{ll}
 20.6 \pm  5.7 ~~~~ {\zbar} \simeq 0.15
 \\
 13.6 \pm  4.3 ~~~~ {\zbar} \simeq 0.35 
\end{array}
\right.
\end{array}
\label{s3}
\eeq

In Fig.\ref{s3edr}
we have used the masked mapped with $0.2$ reddening and $2"$ seeing.
Results for other choices of the mask parameters are similar. 

\section{Comparison with predictions}

Consider the case of power-law correlations: $\xi(r)
=\left({r_0\over{r}}\right)^{\gamma}$ or $P(k) \sim k^n$, with 
$\gamma=n+3$.  In current models of structure
formation $n$ or $\gamma$ vary only smoothly with scale, so 
this should be a good approximation if we limit our study to a small
range of scales. We will normalized the amplitude as:
\beq
\xibar (R) = \sigma_{0.2}^2 \left({0.2 Mpc/h\over{R}} \right)^\gamma
\label{xibar02}
\eeq
where $\sigma^2_{0.2}$ refers to the {\it non-linear} amplitude
of mass fluctuations on scales of $0.2 Mpc/h$, which correspond
to $\sim 1'$ in the sky.

Galaxies might not be fair traces of the underlaying mass distribution.
This is in fact one of the main motivations to use weak lensing as a tool
to study large scale structure.
In the Appendix we give a prescription to parameterize the possible bias
that relates galaxies and matter distributions.  For a power-law correlation,
this parameterization allows for a shift $b_{0.2}$ in the amplitude and
also a shift $ \gamma_b$ in the slope of the galaxy correlation with respect
to that of the matter distribution.

\subsection{Projection and lens magnification}

To simplify notation, and
without lost of generality, we will give all expression for a flat universe,
$\Omega_m+\Omega_\Lambda=1$,
where the comoving angular distance $r(\chi)$ equals the radial
comoving distance $\chi$ (see Bernardeau, Van Waerbeke \& Mellier 1997,
Moessner \& Jain 1998, for the more general case). Also, by default, assume
$\Omega_m = 0.3$, in agreement with current observations (see \S6).

The  $\chi$ distance
is given in terms of the redshift $z$ by:

\beq
d\chi = {dz\over{E(z)}}  ~~~;~~~ E(z)={H_0\over{c}} \sqrt{1-\Omega_m+(1+z)^3 \Omega_m}
\eeq
which can be used to map $\chi=\chi(z)$ and $z=z(\chi)$.
For  $\zbar\simeq 0.15$ we find a mean $\chi \simeq 430$ Mpc/h, while
$\zbar\simeq 0.35$ we have $\chi \simeq 960$ Mpc/h.

The projected galaxy fluctuation is:

\beq
\delta_G(\theta) = \int_0^{\chi_H} ~d\chi ~W_G(\chi) ~\delta_G(\chi,\theta) 
\label{deltag}
\eeq
where $\chi_H$ is the distance to the horizon and
  $\theta$ refers to either the angular position in the  sky or the
radius of a circular cell in the sky. In the later case, which is the
one we will study, $\delta$ is  smoothed over
a cell or radius $\theta$  (a cone in the sky). The galaxy selection function
$W_G(\chi)$ corresponds to the probability of including a galaxy in
the survey and is normalized to unity:

\beq
\int_0^{\chi_H} ~d\chi ~W_G(\chi) = \int_0^{z_H} ~dz ~n_G(z) = 1
\eeq
where $n(z)$ is the normalized redshift distribution.

On the other hand, fluctuations in the flux limited QSO 
induced by weak lensing magnification, $\delta_\mu$, can be expressed in terms
of the weak lensing convergence $\calk$: 

\beq
 \delta_\mu(\theta) = 2 ~(\alpha -1) ~ \calk(\theta) 
\label{deltamu1}
\eeq
where  $\alpha$ is the slope of the QSO number counts:
\beq
N(>i'_Q) \sim 10^{0.4~\alpha~
  i'_Q}
\label{alfa}
\eeq
In our sample we find a least square fit of $\alpha \simeq 1.88 \pm
0.03$ in the range $18.4 < i'_Q < 18.9$ (shown as continuous line 
in Fig.\ref{edrqsoccounts}).\footnote{At fainter $m_i>19$
 magnitudes the counts show  
a very sharp break which could be related to possible incompleteness in our sample.
Thus, can not estimate the slope reliably for the faint QSO sub-sample $18.8<i'_Q<19.1$.}
The convergence is given by a projection over the radial 
{\it matter} fluctuation $ \delta(\chi,\theta)$, that
acts as a lens. This projection is an integral over the lensing magnification efficiency,
$\cale(\chi)$, a geometrical factor $\simeq {{\chi_Q-\chi}\over{\chi_Q}}$
 that depends on the QSO $\chi_Q$ radial distribution and $\Omega_m$.
We express this projection as:
\beq
\delta_\mu(\theta) = \int_0^{\chi_H} ~d\chi ~\cale(\chi) ~\delta(\chi,\theta) 
\label{deltamu2}
\eeq
where
\beq
\cale(\chi) = 2(\alpha -1){3 H_0^2\Omega_m\over{2~c^2}}
(1+z) \chi \int_\chi^{\chi_H} d\chi'
 {{\chi'-\chi}\over{\chi'}}~ W_Q(\chi')
\eeq
and $W_Q(\chi)$ is given by the normalized probability to include a QSO
in our sample:
\beq
\int_0^{\chi_H} ~d\chi ~W_Q(\chi) = \int_0^{z_H} ~dz ~n_Q(z) = 1
\eeq

\begin{figure*} 
\centerline{\epsfysize=8truecm
\epsfbox{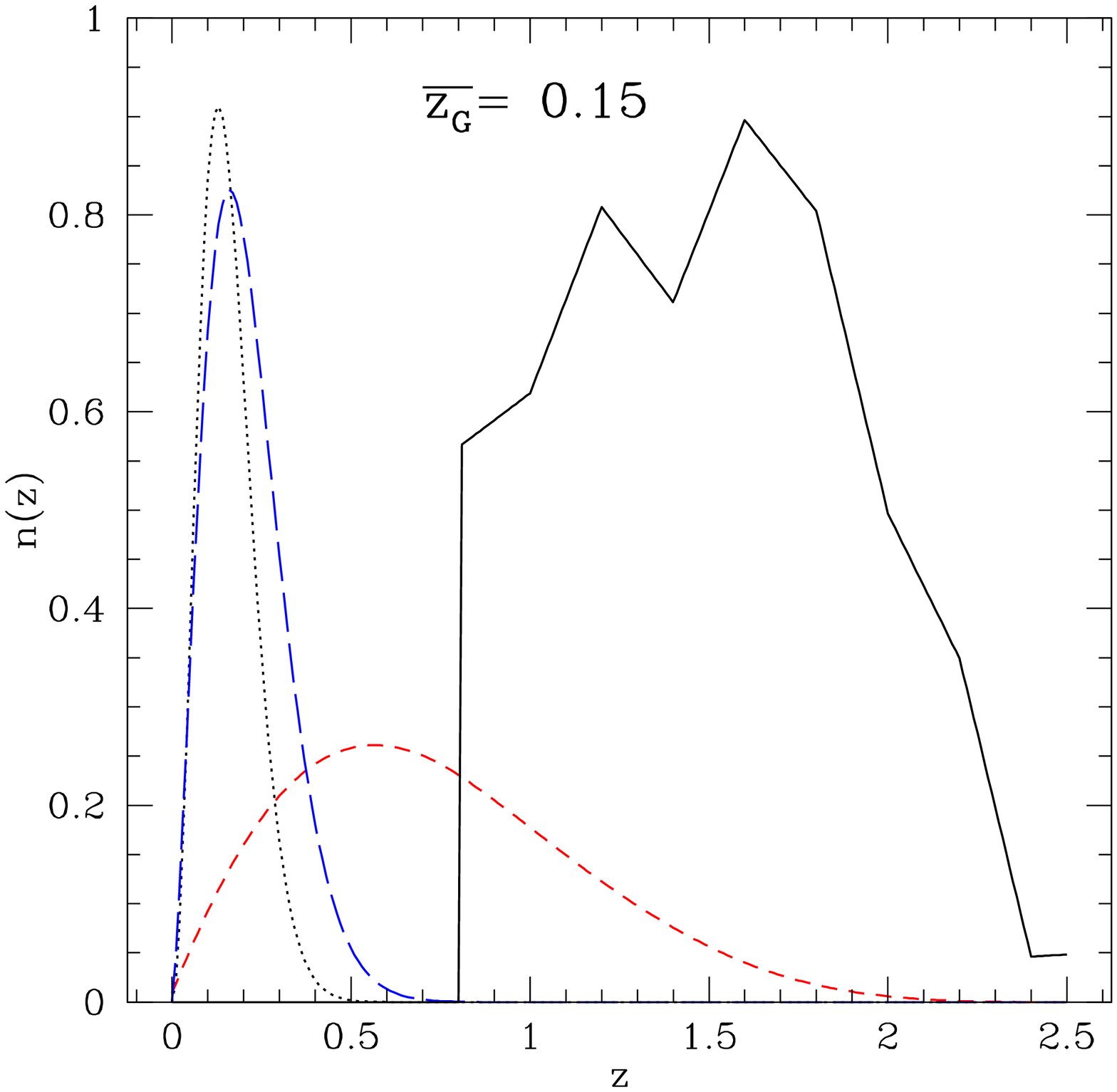}\epsfysize=8truecm
\epsfbox{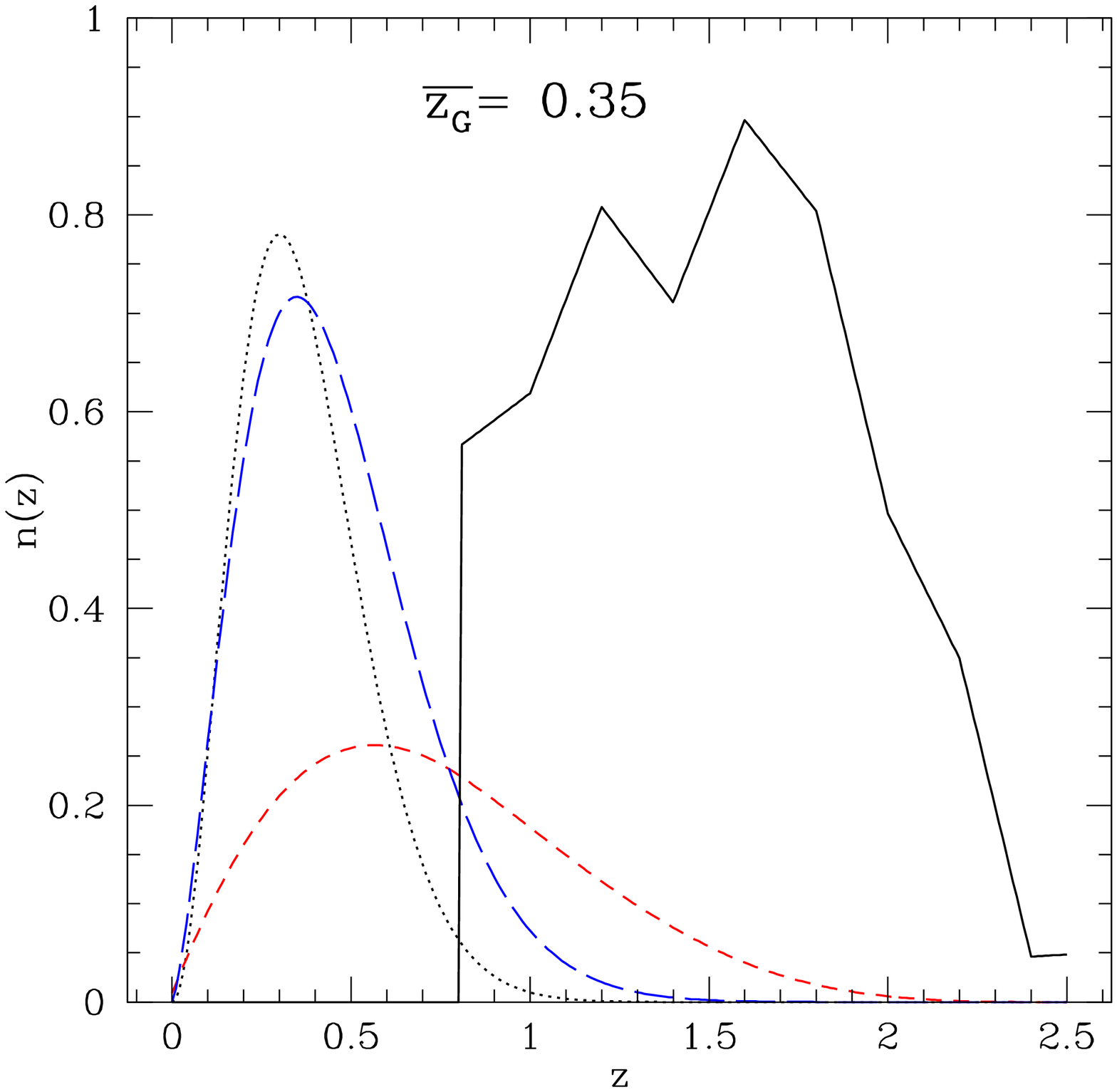}}
\caption[Fig2]{\label{effz} 
The dotted line shows the predicted  galaxy redshift distribution $n_G(z)$ (divided
by 3 to be on scale) for $\zbar= 0.35$ (right panel), 
which corresponds to $r'<22$ (and also $i'<22$) and for $\zbar = 0.15$ (left panel), 
which corresponds to $r'<19$. The continuous line shows the normalized
QSO redshift distribution $n_Q(z)$ as measured in EDR/QSO. 
Short-dashed line is the lensing
magnification efficiency $\cale(z)/E(z)$ for the shown $n_Q(z)$.
The long-dashed line corresponds to $\sqrt{n_G(z)~\cale(z)/E(z)}$ 
which roughly represents the efficiency of galaxy-QSO cross-correlation.}
\end{figure*}

As noted in Bernardeau, Van Waerbeke \& Mellier (1997), the crucial difference
between  galaxy projection in  Eq.[\ref{deltag}] and lensing magnification in
Eq.[\ref{deltamu2}]  is that $\cale$ is not
normalized to unity. Thus, besides the projection effect, which can be modeled
as an stochastic selection function, we have an overall
re-scaling of fluctuation amplitudes (this has been used in Gazta\~naga \&
Bernardeau 1998 to produce simple non-linear weak-lensing simulations). 
Figure \ref{effz} compares $\cale(z)/E(z)$ with $n_G(z)$ and $n_Q(z)$ for the QSO
and galaxy samples in our analysis. For QSOs the redshift distribution is the
one measured in the EDR/QSO $16.5<i_Q<18.8$ sample, while for galaxies we
show the predictions in Dodelson \etal (2001).

\subsection{Mass-mass and galaxy-galaxy correlations}

Using Eq.[\ref{xibar02}]  and within the small
angle approximation( eg see \S 7.2.1 in Bernardeau \etal 2002), 
the variance of the projected {\it mass} fluctuations can be expressed as:
\beq
w(\theta) \equiv < \delta^2(\theta) > =
\sigma_{0.2}^2~ A ~\theta^{1-\gamma} \label{w_2} 
\eeq
where
\bea
A &=& A(\gamma) = B(\gamma)~T_\gamma ~ \overline{W^2_G}      \label{A}
 \\ \nonumber
B(\gamma) & \equiv & {(3-\gamma)~(4-\gamma)~(6-\gamma)~
\Gamma(\gamma/2-1/2)\Gamma(1/2)\over{5^\gamma~2^{3-\gamma}~9~\Gamma(\gamma/2)}}
\eea
$T_\gamma$ is a geometrical
factor of order unity ($T_\gamma \simeq 0.7-0.8$)
that comes from the area average over the 2-point function:
\beq
T_\gamma \equiv  {4\over{\pi(5-\gamma)}} \int_0^1 x~dx 
\int_0^{2\pi} d\phi \left(1+x^2-2x\cos{\phi}
 \right)^{{1-\gamma}\over{2}},
\eeq
and $\overline{W^2_G}$ is given by the radial galaxy selection:
\beq
\overline{W^2_G} \equiv \int_0^{\chi_H} ~d\chi ~W_G^2(\chi) ~\chi^{1-\gamma} D^2(z)
\label{w2G}
\eeq
where  $D(z)$ accounts for the redshift evolution of the correlation function:
eg in the linear regime $D(z)$ is the linear growth factor, in the stable
clustering regime $D^2(z)=(1+z)^{-3}$.

For the galaxy-galaxy variance we assume the bias in Eq.[\ref{bgamma_b}],
which   allows for a shift $b_{0.2}$ in the amplitude and
also a shift $ \gamma_b$ in the slope of the galaxy correlation with respect
to that of the matter distribution. Thus, we have:
\beq
w_{GG}(\theta) \equiv < \delta^2_G(\theta) > = 
b_{0.2}^2 ~\sigma_{0.2}^2~ A_{GG}~\theta^{1-\gamma-2\gamma_b} 
\label{wGG}
\eeq
where $A_{GG}$ is the same as $A$ in Eq.[\ref{A}] for the new slope:
$A_{GG}= A(\gamma+2\gamma_b)$.

\subsection{QSO-QSO and galaxy-QSO correlations}

We next want to estimate
the galaxy-QSO cross-correlation. We express the angular QSO
fluctuations as:
\beq
\delta_Q(\theta) = \delta_Q^I(\theta) +  \delta_\mu(\theta)
\eeq
where $\delta_Q^I(\theta)$ stands for the intrinsic fluctuations
while $\delta_\mu(\theta)$ are fluctuations induced by magnification
bias. Neglecting the QSO (source) and matter (lens) cross-correlation
(which is negligible given the large radial separations),
the observed QSO variance has two contributions:
\beq
w_{QQ}(\theta) \equiv < \delta_Q(\theta)~\delta_Q(\theta)> =
w^I_{QQ}(\theta) + w_{\mu\mu}(\theta)
\eeq
where:
\beq
w_{\mu\mu}(\theta) \equiv  < \delta^2_\mu(\theta)> 
= \sigma_{0.2}^2~ A_{QQ}~\theta^{1-\gamma}
\eeq
with:
\bea
A_{QQ} &=& B(\gamma)~T_\gamma ~\overline{\cale^2} \\ \nonumber
\overline{\cale^2} &\equiv & \int_0^{\chi_H} ~d\chi 
~\cale^2(\chi) ~\chi^{1-\gamma} D^2(z).
\eea
The above expressions can be used, for a known magnification, 
to separate the intrinsic from the apparent QSO clustering.

The QSO and galaxy populations are well separated in radial distances
so that 
we can neglect the intrinsic
cross-correlation $<\delta_G(\theta)~\delta_Q^I(\theta)>$. We then have:
\beq
w_{GQ}(\theta) \equiv < \delta_G(\theta)~\delta_Q(\theta)> \simeq
< \delta_G(\theta)~\delta_\mu(\theta)>
\eeq
so that using Eq.[\ref{deltamu2}]  and assuming Eq.[\ref{bgamma_b}]
we find:
\beq
w_{GQ}(\theta) = b_{0.2} ~\sigma_{0.2}^2~ A_{GQ}~\theta^{1-\gamma-\gamma_b}, 
\eeq
with:
\bea
A_{GQ} &=& B(\gamma+\gamma_b)~T_{\gamma+\gamma_b} ~\overline{W_G \cale} \\ \nonumber
\overline{W_G \cale} &\equiv &  \int_0^{\chi_H} ~d\chi ~W_G(\chi)~\cale(\chi) ~\chi^{1-\gamma-\gamma_b} D^2(z)
\eea

\subsection{Measure of bias and matter fluctuations}

The combination galaxy-galaxy and galaxy-QSO 
cross-correlation will allow us to break the
intrinsic degeneracy between biasing and matter fluctuations, eg
between $b_{0.2}$ and $\sigma_{0.2}$ as measured by galaxy surveys alone.
Here, we propose to measure the four parameters that characterize
bias and matter fluctuations in a narrow range of scales (the
power-law approximation). These parameters are $\sigma_{0.2}^2$ and $\gamma$
for the variance in non-linear {\it mass} fluctuations (eg Eq.[\ref{xibar02}])
and  $b_{0.2}$, $\gamma_b$ for the bias function as described in the Appendix, eg
 Eq.[\ref{bgamma_b}].

We can estimate
these four parameters, $\sigma_{0.2}$, $b_{0.2}$, $\gamma$ and $\gamma_b$,
from the angular observations of galaxy-QSO
correlations in the following way.
We first take the measured  logarithmic slopes of $w_{GG}$ and
$w_{GQ}$ in Eq.[\ref{betaGG}]-[\ref{betaGQ}] 
from a fit to the data in Fig.\ref{w2gqfit}. These slopes and the used
 to find the intrinsic {\it matter} slope $\gamma$ and 
bias scale dependence $\gamma_b$:

\bea
\gamma &=& 1+ \beta_{GG} - 2 \beta_{GQ}
\nonumber \\ \label{gammab}
\gamma_b &=& \beta_{GQ} - \beta_{GG}
\eea

 We can then obtained $b_{0.2}$ and $\sigma_{0.2}$ as:
\bea
b_{0.2} &=& {{w_{GG}(\theta)}\over{w_{GQ}(\theta)}} ~  {A_{GQ}\over{
    A_{GG}}}  ~\theta^{\gamma_b} \nonumber \\ \label{bs02}
\sigma^2_{0.2} &=&  {{w_{GQ}^2(\theta)}\over{w_{GG}(\theta)}} ~
 {A_{GG}\over{
    A^2_{GQ}}}  ~\theta^{\gamma-1}
\eea
If the power-law model is a good approximation these values should not be
a strong function of scale. This will provide a consistency test for 
our power-law approximation.

The above reconstruction scheme has been tested
against mock angular simulations in Gazta\~naga (1995). 
This method can be generalized to extract the shape $\xibar(R)$ from
$w_{GG}(\theta)$ in the quasi-scale invariance approximation by performing
local inversions around $R \simeq \theta \chi$, where $\chi$ is the
mean depth of the sample (see Gazta\~naga 1995 for details).
We have also verified the validity of such approximations over the
weak lensing simulations presented in Gazta\~naga \& Bernardeau (1998).

\begin{figure*} 

\centerline{\epsfysize=8truecm
\epsfbox{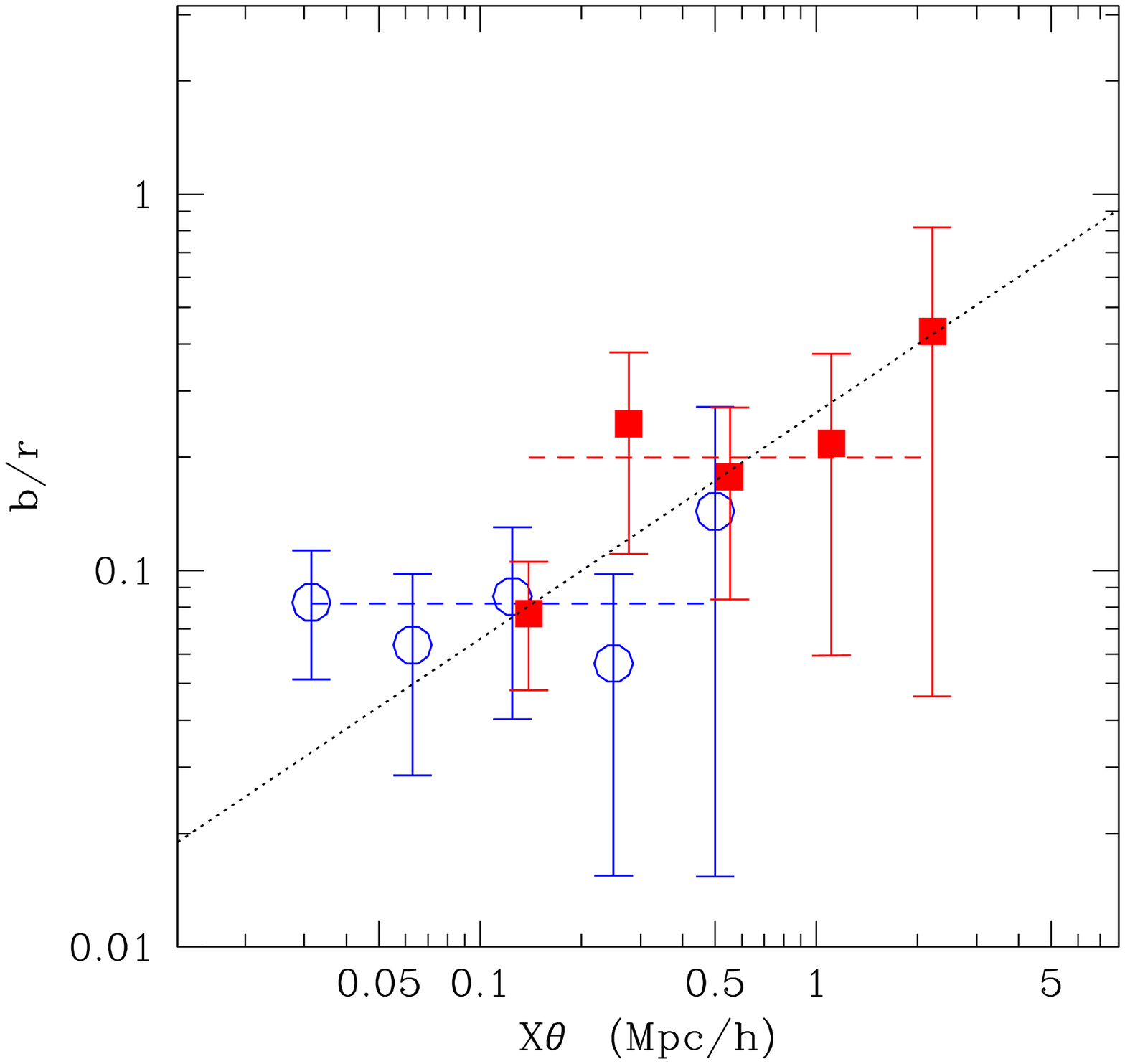}\epsfysize=8truecm
\epsfbox{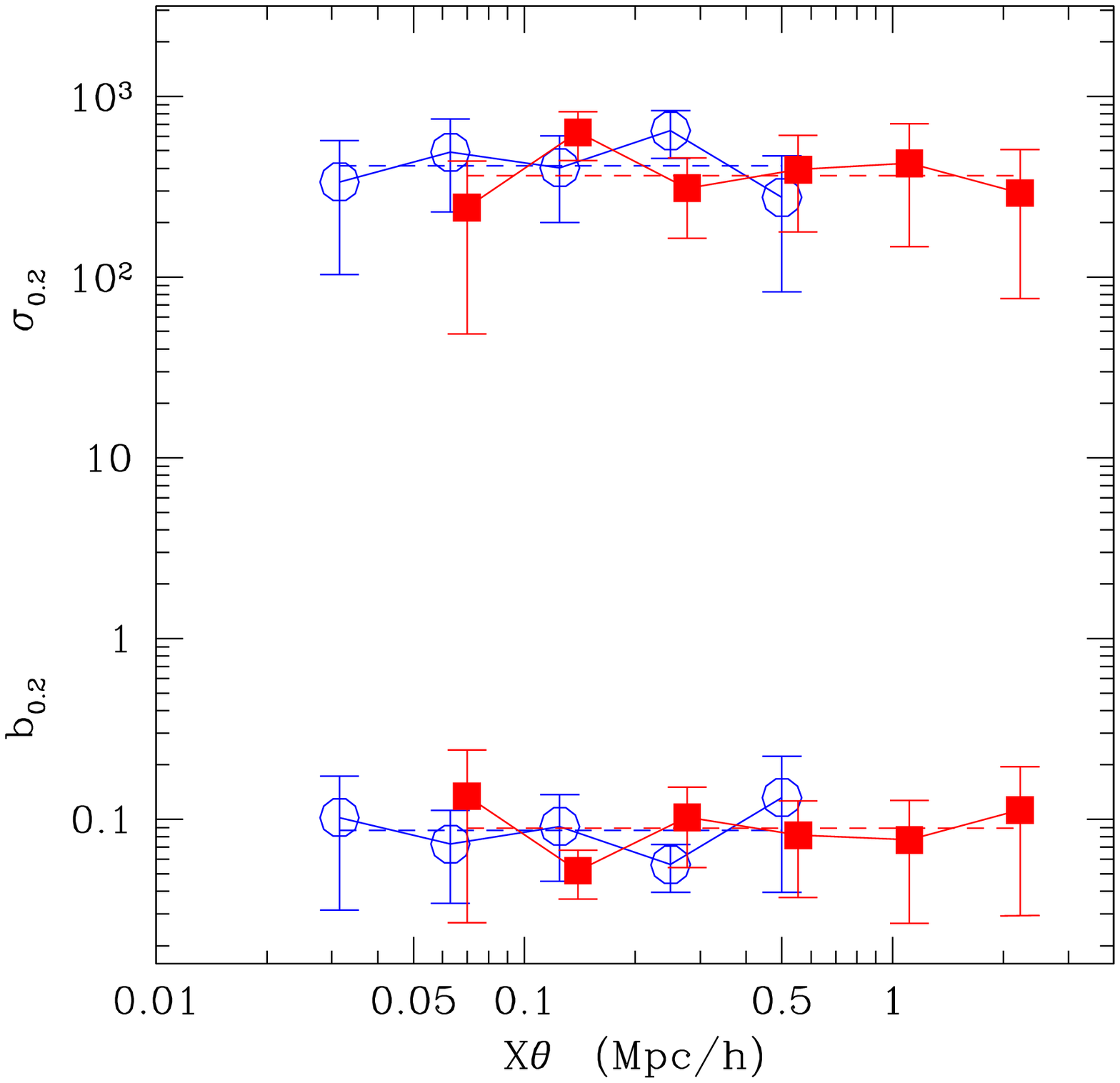}}
\caption[Fig2]{\label{w2bover} 
In both panels: open circles correspond to $r'<19$ ($\zbar=0.15$)
and closed squares to  $i'<22$ ($\zbar=0.35$).
{\sc Left Panel:} Values of $b/r$ de-projected from the ratio $w_{GG}/w_{GQ}$
assuming a constant, but stochastic, bias. Dashed lines are the best fit
constant value of $b/r$ for each galaxy sample. The dotted line gives the best
fit linear relation for both samples.
{\sc Right Panel:} Values of $b_{0.2}$ and $\sigma_{0.2}$ de-projected using
Eq.[\ref{bs02}], as a function of scale.}
\end{figure*}

\subsection{ Scale dependence bias}

Left panel of Fig.\ref{w2bover} shows $b/r$ defined in Eq.[\ref{roverb}] 
and de-projected from the data at each point as 
\beq
{b\over{r}} = {w_{GG} \over{w_{GQ}}}
{A_{GQ}\over{A_{GG}}}, 
\eeq
ie assuming that $b/r$ is scale independent (ie $\gamma_b=0$). Each point is
shown at scale corresponding to the mean depth in the sample $\chi \theta$. 
The resulting values of $b/r$ show a tendency to increase with scale, which
means that $\gamma_b=0$ is not such a good approximation.
But note that this trend is not very significant given the large errorbars: 
a constant value of $b/r$ is only ruled out with a 3-sigma significance.

Right panel of Fig.\ref{w2bover} shows the recovered
values of $\sigma_{0.2}$ and $b_{0.2}$ as a function of scale given the
prescription in Eq.[\ref{bs02}]. 
The recovered values are quite flat,
in good agreement with the assumption that the correlation functions
can be approximated locally by power-laws (eg Eq.[\ref{wGG}]).

\subsection{Results for $\sigma_{0.2}$, $b_{0.2}$, $\gamma$ and $\gamma_b$}

From  Eq.[\ref{betaGG}]-[\ref{betaGQ}] and Eq.[\ref{gammab}]:

\beq
\begin{array}{ll}
 \gamma  ~=
 & \left\{
\begin{array}{ll}
 1.94 \pm  0.34 ~~~~ \zbar \simeq 0.15
 \\
  2.33 \pm  0.26 ~~~~ \zbar \simeq 0.35 
\label{gamma}
\end{array}
\right.
\end{array}
\eeq

\beq
\begin{array}{ll}
 \gamma_b  ~=
 & \left\{
\begin{array}{ll}
 -0.11 \pm 0.17 ~~~~ \zbar \simeq 0.15
 \\
 -0.36 \pm 0.13    ~~~~ \zbar \simeq 0.35 
\label{gammab2}
\end{array}
\right.
\end{array}
\eeq

While the mean values from right panel of Fig.\ref{w2bover} are:

\beq
\begin{array}{ll}
 b_{0.2} ~=
 & \left\{
\begin{array}{ll}
 0.09  \pm 0.02 ~~~~ \zbar \simeq 0.15
 \\
 0.09  \pm 0.02 ~~~~ \zbar \simeq 0.35 
\end{array}
\right.
\end{array}
\eeq

\beq
\begin{array}{ll}
 \sigma_{0.2}  ~=
 & \left\{
\begin{array}{ll}
 412 \pm 97    ~~~~ \zbar \simeq 0.15
 \\
 364 \pm 86    ~~~~ \zbar \simeq 0.35 
\end{array}
\right.
\end{array}
\eeq

These values assume that clustering is fixed in comoving coordinates. 
Values for other clustering evolution, 
ie $\epsilon \ne 0$, can be  obtained 
using in Eq.[\ref{w2G}]:
\beq
D^2(z) \simeq (1+z)^{-(3+\epsilon)}
\eeq
If we require the underlaying ($z=0$)
$\sigma_{0.2}$ to be the same in both samples we find:
\beq
\epsilon \simeq 0.8 \pm 1.3
\label{epsilon}
\eeq
which roughly agrees with stable clustering, but has a large error.

\section{Discussion}

As we will show in some detailed here, the values we have recovered for
$\sigma_{0.2}$, $b_{0.2}$, $\gamma$ and $\gamma_b$ appear in contradiction
with currently popular expectations.  We should therefore discuss first the
two more important systematics in our analysis: error-bar estimation (section
\S6.1) and extinction (\S6.2).

In \S6.3 we give a brief account of what are the currently popular
expectations for the above observables. In \S6.4 we will show how
the results we find in this paper seem to contradict the
expectations. A possible solution for this contradiction is hinted
in \S6.5 and \S6.6 as we consider clustering evolution 
in shape and amplitude. In subsection 6.7 we discuss how our
interpretation changes with $\Omega_m$ and find that
 our proposed solution can also explained the low value
of the measured skewness. Finally in \S6.8, we compare our
results with other estimates.

\subsection{Unrealistic or correlated error-bars?}

We have sliced the data, as displaced in Fig.\ref{pixelMap}, in $N=10-40$ RA
bins (horizontal direction in the Figure).  Because of the mask, this results
in subsamples with different shapes and holes, which increases the
sample-to-sample scatter in the jackknife error estimation Eq[\ref{jack}].
Thus we believe that this approach is conservative
and slightly overestimates all the  jackknife errorbars presented in the
above section. We have tested this idea with the
APM (Maddox \etal 1990) pixel maps.

 We compare the jackknife error over a given slice out of the APM map
against the scatter from different slices (the APM can fit over 20 EDR
slices). The jackknife error is comparable to the zone to zone scatter on
scales larger than a few arc-minutes (where the mean density of galaxies in a
cell is larger than unity), but is about a factor of 3 larger than the zone to
zone errors on the smallest scales (dominated by shot-noise fluctuations).  We
do not attempt to correct for this here, but rather take the conservative approach
of using the jackknife errors. This means that in principle the significance of
the analysis presented below can be improved with a more sophisticated
treatment of errors (eg see Colombi, Szapudi \& Szalay 1998).

There is also an overall shift in the mean amplitude of the
zone to zone correlations (variance or skewness), due to
large scale (sample size) density fluctuations, which is not taken into
account by the jackknife error. The amplitude of this effect is comparable to
the jackknife error on 20' scales. This is similar to an overall calibration
error as indeed corresponds to the uncertainty in the value of the mean
density over the whole map (eg see Hui \& Gazta\~naga 1999).  
Our error analysis hardly changes if we take this into account
as we are mostly dominated by the (relatively large) scatter
from shot-noise and small scale fluctuations.

Other than this, the covariance shown
in bottom-right panel of  Figure \ref{w2mz} seems to be dominated by
the diagonal terms and we have neglected it. 

\subsection{Variable obscuration from the Galaxy}

Could our result be caused by small scale variations from obscuration in our
galaxy? In \S 3 we have shown that large scale extinction is not affecting our
results. There is a small but significant anti-correlation of the SDSS/EDR
galaxy and QSO maps with Schlegel \etal (1998) which induces an artificial
galaxy-QSO cross-correlation. We have shown how this effect disappears when
using an extinction mask. But Schlegel \etal (1998) maps have a FWHM of
$\theta=6'$, while we measure correlations below $\theta=1'$. Could our
cross-correlation be caused by extinction at smaller scales?  The spectrum of
extinction fluctuations as measured by Schlegel \etal (1998), $P(k) \propto
k^{-2.5}$ at scales from $\simeq 6$ arc-min to $\simeq 30$ degrees. ,
indicates that the amplitude of density fluctuations induced by absorption
$<\delta^2_A> \propto \theta^{+0.5}$ grows slowly with scales.  Thus
extinction produces larger galaxy-QSO artifacts on larger scales.

Extrapolating this spectrum to smaller scales should produce smaller
contributions than the ones we have already estimated in \S 3.\footnote{In a recent
paper, Kiss \etal (2002) find galactic cirrus emission to varie from field to
field as $P(k) \propto k^\alpha$ with $-5.3<\alpha <-2.1$ and
even stipper spectrum at smaller scales. The extreme case $\alpha\simeq 2$
produces a constant value of $<\delta^2_A>$ as a function of scale, which can 
not account for our observations.}
 In summary, the
detected galaxy-QSO cross-correlation have therefore the wrong amplitude and
the wrong shape to be explained by galactic extinction.

\begin{figure*} 
\centerline{\epsfysize=8truecm
\epsfbox{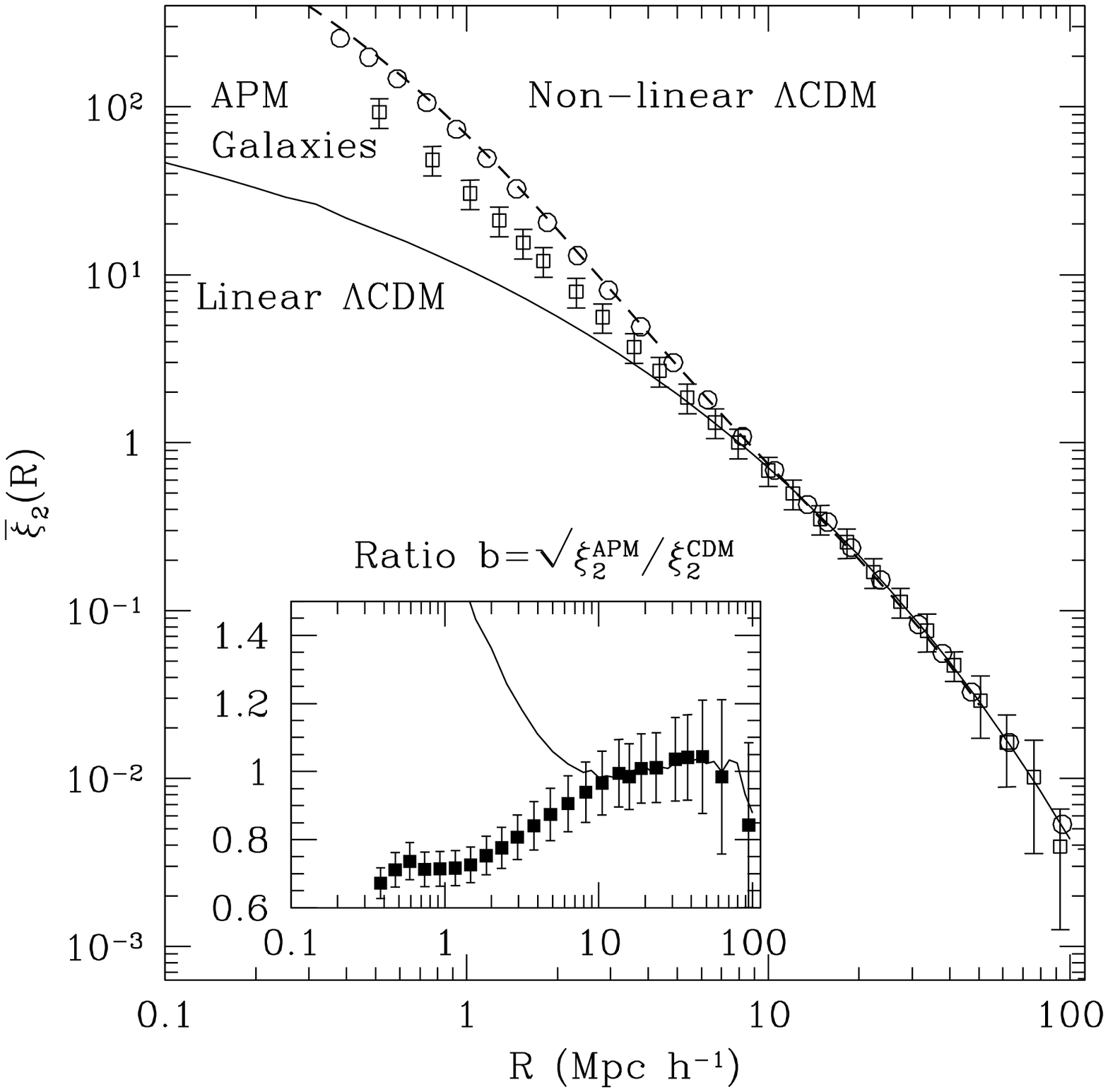}\epsfysize=8truecm
\epsfbox{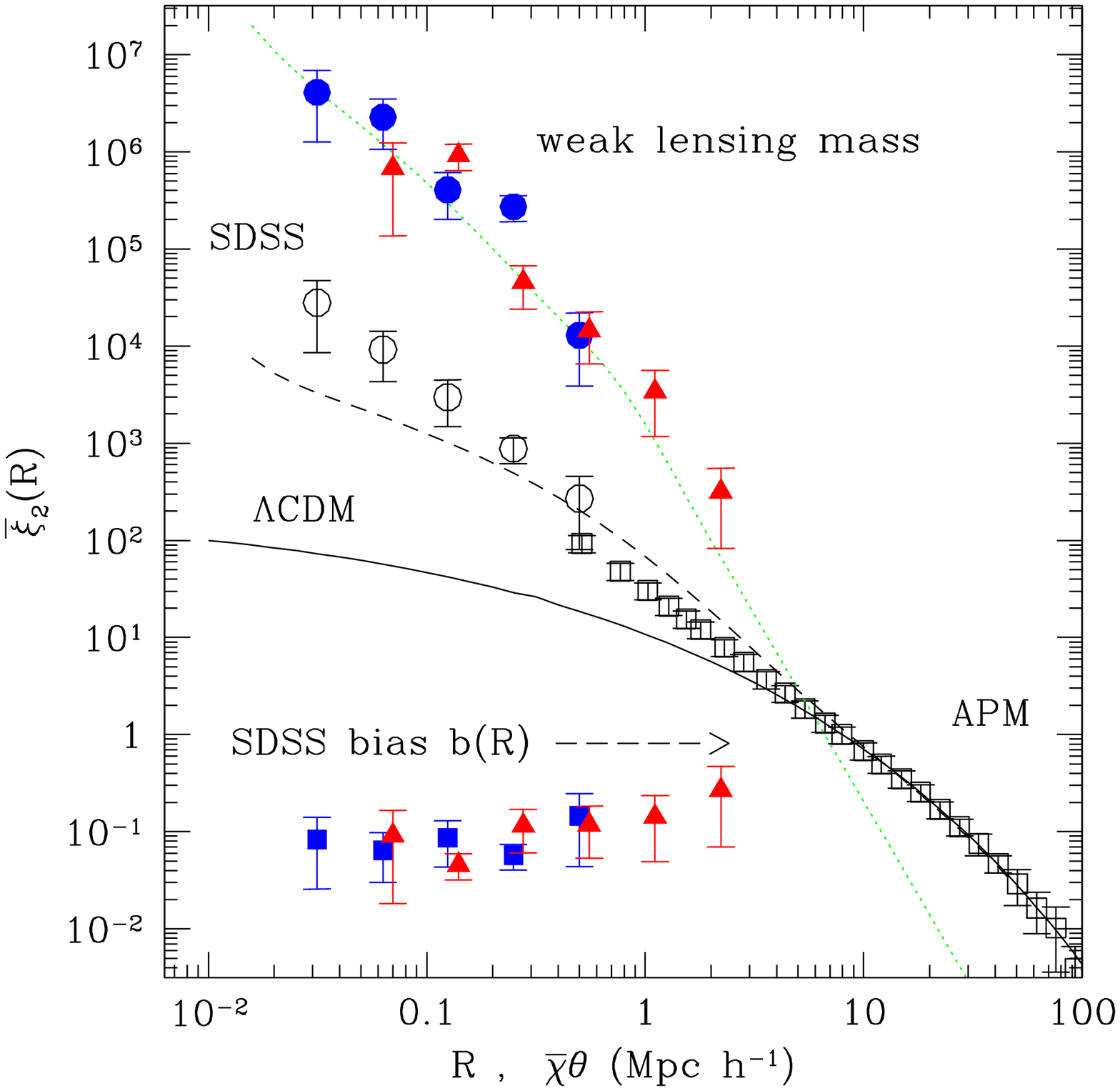}}
\caption[Fig2]{\label{x2c80sdss}
{\sc Left panel:}
Comparison of the linear (continuous line) and non-linear (open squares)
variance in the $\Lambda$CDM simulations ($\xibar^{\Lambda CDM}$)
with that in  APM 
galaxies ($\xibar^APM$, squares with error-bars), from Fig.11 in Gazta\~naga (1995).
Dashed lines shows the non-linear fitting formulae of Smith \etal (2002).
The inset shows the bias function $b(R)\equiv
\sqrt{{\xibar^APM\over{\xibar^{\Lambda CDM}}}}$ for the non-linear (closed squares with
errorbars) and linear  $\Lambda$CDM (continuous line) cases.
{\sc Right panel:}
Comparison of the linear (continuous line) and non-linear (dashed line)
variance in the $\Lambda$CDM model
with that from the $b_J <20$ APM 
galaxies (open squares with error-bars) and $r'<19$ SDSS galaxies (
open circles with error-bars). Upper closed circles and triangles show
the non-linear variance  reconstructed from the galaxy-QSO
correlation in the EDR/SDSS at $\zbar \simeq 0.15$ and $\zbar \simeq 0.35$
(scaled to $z=0$ under stable clustering). Lower closed figures show the
corresponding results for the bias function $b(R)$.
The dotted line shows the non-linear mass variance prediction 
for a steep $n_{eff} \simeq 0.97$
initial spectrum.}
\end{figure*}

\subsection{The standard  $\Lambda$CDM picture}

In the non-linear regime dark-matter ($\Lambda CDM$) numerical simulations
typically find that on cluster scales ($r \simeq 0.1$ Mpc/h) the non-linear
variance of density fluctuations is $\xibar_{NL} \simeq 10^2 \xibar_L$, where
$\xibar_L$ stands for the linear value (compare continuous line with
short-dashed line in left panel of Fig.\ref{x2c80sdss}). Moreover $\xibar_{NL}
\sim R^{-1}$ ( $\sim R^{-3/2}$ at the non-linear transition) which is
typically steeper than the linear values $\xibar_{L} \sim R^{-1/2}$ (
$\sim R^{-1}$ at the non-linear transition).  This behavior is reproduced by
the non-linear fitting formulaes (eg Peacock \& Dodds 1996) and have been
extensively used in weak-lensing predictions (eg Bartelmann \& Schneider 2001,
Benitez \etal 2001).  For galaxies, predictions vary from model to
model but one can typically find that on non-linear scales (1 to 10 Mpc/h)
$\xibar_G \simeq \xibar_L$ for blue
(Late-type) galaxies and up to $\xibar_G \simeq 10 \xibar_{NL}$ for the red
(Early-type) population.  This gives a range $b \simeq
\sqrt{\xibar_G/\xibar_{NL}} \simeq 0.1$ and $b \simeq 3$.  This is also in
rough agreement with the estimations in galaxy surveys.  We will call this the
{\it standard $\Lambda$CDM picture} for non-linear galaxy and mass evolution
(see Cooray \& Sheth 2002 for an excellent review on all the above ideas).

In terms of the variance most of these features can be summarized in the left
panel of Figure \ref{x2c80sdss}, which is a expanded version of Fig.11 in
Gazta\~naga (1995). Note how APM galaxies seem to be anti-bias on small scales
with respect the $\Lambda$CDM model, something that seems to be compatible
with different models of galaxy formation (eg Jenkins \etal 1998, Seljak 2000,
Scoccimarro \etal 2001a, Sheth \etal 2001 and references therein).

\subsection{Weak lensing magnification}

If we take the detected galaxy-QSO cross-correlation signal as being purely
due to weak lensing magnification, we find important challenges for the above
picture, even in the orders of magnitude involved. These are summarized in
right panel of Fig.\ref{x2c80sdss}, which shows the recovered mass variance as
a function of scale.  We find $b \simeq 0.1$ (which seems to agree with the
above blue galaxy model above) but note that $\xibar_{NL} \simeq 10^4
\xibar_L$ (ie compared upper closed figures, $\xibar_{NL}$, with dashed line,
$\xibar_L$) which is much larger than expected in the {\it
  standard $\Lambda$CDM picture} above.

We also find some evolutionary trends.  First of all, from Eq.[\ref{gamma}]
we find $\xibar_{NL}
\sim R^{-2}$ which is steeper (3 sigma significance) than in the {\it
  standard $\Lambda$CDM picture} (slope of the dashed line in right panel
of Fig.\ref{x2c80sdss}).  At higher redshifts ($\zbar \sim 0.35$) the
slope seems slightly steeper $\xibar_{NL} \sim R^{-2.3}$, as expected if we
assume that on average there are lower mass halos at higher redshifts, and
therefore steeper profiles (see Navarro, Frenk \& White 1996).  Obviously, the
slope of the galaxy $\xibar_G$ distribution is the same as in the {\it
  standard $\Lambda$CDM picture} (as these are direct observations that we
also reproduce in our analysis).

The measured values of the amplitude today $\sigma_{0.2} \simeq 400$ is about
$100$ times larger that expected in the {\it standard $\Lambda$CDM picture},
although the discrepancy factor depends on the particular cosmology.  The
significance of this result is over 4-sigma.

The biasing amplitude $b$ at $0.2$ Mpc/h remains constant $b\simeq 0.1$ with
redshift.  The null hypothesis of $b=1$ is ruled out at a large significance.
The slope $\gamma_b$  (ie from Eq.[\ref{gammab2}])
seems to become more negative at $\zbar \sim 0.35$ (only
1-sigma effect), which partially masks the steepening of the density profiles
as we increase the redshift.

On weakly non-linear scales, the shape of the galaxy 3-point function and
bispectrum can be used to determine $b$ with independence of the underlaying
spectrum. This was first proposed by Frieman \& Gazta\~naga (1994) and Fry
(1994), following the ideas in Fry \& Gazta\~naga (1993). Recent measurements
( Frieman \& Gazta\~naga 1999, Scoccimarro \etal 2001b, Verde et. al 2002) find
$b\simeq 1$ at scales $> 10$ Mpc/h. In fact this value is not in contradiction
with the values above as the bias seems to increase with scale, which
extrapolates well to $b\simeq 1 $ at $\simeq 10$ Mpc/h.

Our recovered valued do not assume any model for the primordial spectrum or
its subsequent evolution, but assumed a flat geometry with $\Omega_m=0.3$.
This affects both the projection and the amplitude of the lensing
magnification.  The geometry enters in both the lensing and the galaxy
projections and should therefore cancel out to some extend for $b$, but not
quite for $\xibar_{NL}$ which is estimated from ratio $w^2_{GQ}/w_{GG}$. Thus
we expect our estimations of $b$ to scale with $\sim {\Omega_m\over{0.3}}$
and $\xibar_{NL}$ as $\sim ({0.3\over{\Omega_m}})^{1.5}$ (see Bernardeau
\etal 1997).  Thus if $\Omega_m \simeq 1$ this would produce $b \simeq 0.3$
and $\sigma_{0.2} \simeq 158$. These values are closer to the {\it standard
  $\Lambda$CDM picture} above but still uncomfortably far from it. Moreover
they represent an inconsistency in the value of $\Omega_m$.

If the {\it standard  $\Lambda$CDM picture} fails on these small scales, is there
any possible alternative?

\subsection{Stable clustering}
\label{sec:stable}

 The amplitude of $\xibar_{NL}$ at $0.2$ Mpc/h follows
the stable clustering regime (structures are decoupled from the Hubble
expansion): $\xibar \sim (1+z)^{-(3+\epsilon)}$, 
with $\epsilon \simeq 0$,
indicating that indeed clustering
could be dominated by the halo profile ($1h$) rather than by the relative
distribution of halo pairs ($2h$ term in Eq.[86] of Cooray \& Sheth 2002). 
The fact that the recovered amplitudes at redshifts
$\zbar_1\simeq 0.15$  and $\zbar_2\simeq 0.35$ 
follow:
\beq
{\xibar(\zbar_1)\over{\xibar(\zbar_2)}} \simeq
\left({1+\zbar_1\over{1+\zbar_2}}\right)^{-3} \simeq 1.6
\eeq
does not involved any parameter fitting and 
is non trivial given  that the raw
observed correlation at different redshifts are quite different (eg see
Fig.[\ref{w2gqfit}]). In other words, there is no reason to expect any
systematic effect to mimic this redshift dependence.  Note that both in the linear
regime $\xibar_L \simeq D^2 \xibar_0$ and in the strongly non-linear
regime, this ratio should be independent of the amplitude $\xibar_0$ of the
primordial spectrum. The fact that is good agreement with stable
clustering on scales where clusters have collapsed is not surprising.
Unfortunately the error we find for $\epsilon$ in Eq.[\ref{epsilon}]
is too large to draw more significant conclusions.

There is another aspect to stable clustering, which is related to the scale
dependence  of the correlations. Peebles (1980) have shown that 
in the stable clustering for
a  power-law initial spectrum $\xibar_{L} \sim R^{-(n+3)}$
(where $n$ is the index of the linear spectrum  $P(k) \sim k^n$), the
non-linear slope $\xibar_{NL} \sim R^{-\gamma}$ is:
\beq
\gamma = {3(n+3)\over{5+n}}
\label{stable}
\eeq
In the  $\Lambda$CDM models $n \simeq -1.5$ on non-linear scales so that one
would expect $\gamma \simeq 1.3$.
Smith \etal find that N-body simulations in fact fall a bit short 
of this prediction, with $\gamma \simeq 1.0$. For other values of
$n$ they find that $\gamma$ increases with the steepness roughly
as predicted by the stable clustering, but the measure $\gamma$ always 
falls a bit below the stable clustering prediction in Eq.[\ref{stable}]
(see Fig. 9 in Smith \etal 2002).
It should be noted that the agreement seems to improve for steep
slopes: ie at $n=0$, 
 Smith \etal (2002) find $\gamma \simeq 1.7$ while Eq.[\ref{stable}]
gives $\gamma = 1.8$. Also note that Eq.[\ref{stable}] might be valid
for stronger non-linearities (or smaller scales) than tested in the
Smith et al (2002) simulations.

Thus, the value we find from the galaxy-QSO data, $\gamma \simeq 2$ 
in Eq.\ref{gamma}], indicates
a spectrum $n \simeq 1$ in Eq.[\ref{stable}]. This value is very different
from the {\it standard $\Lambda$CDM picture}.  But we will now show that in
fact, for such steep initial spectrum normalized to $\sigma_8 \simeq 1$, one
expects the non-linear amplitude to be $\sigma_{0.2} \simeq 400$, just as
found in the our interpretation of the galaxy-QSO correlations.

\subsection{Non-linearities and halo model}

One of the possible origins of the discrepancy that we find for the
measured value of $\sigma_{0.2}$ as compared  with the {\it
standard  $\Lambda$CDM picture} is the modeling of the non-linear
variance $\xibar_{NL}$. Based in the stable clustering ideas, Hamilton et al
(1991) proposed that the non-linear variance should be a fixed universal function
$f_{NL}$ of the linear variance once we identify the adequate mapping of linear
to non-linear scales: $\xibar_{NL} \simeq f_{NL}(\xibar_L)$. Under stable
clustering, the cosmic evolution of  $\xibar_{NL}$ with the scale factor $a$ is
$\xibar_{NL} \sim a^3$ while $\xibar_{L} \sim a^2$, which implies
that $\xibar_{NL} \simeq \xibar_{L}^{3/2}$. These transformations were
generalized to the power spectrum and to models with $\Omega \ne 1$ by Peacock
\& Dodds (1994, 1996) and Jain, Mo \& White (1995) to provide phenomenological
fitting formulaes for non-linear clustering which have since been widely used.
The range of validity and the accuracy of these initial ansatze was 
limited and, when looked in detail, there appears to be a
fundamental problem in getting a universal fit that works for scale dependent
models (see Baugh \& Gazta\~naga 1996, Smith \etal 2002).

A more recent realization of this idea involves the dark matter halo approach
(see Cooray \& Sheth 2002 for a review) which assumes that the density field
can be modeled by a distribution of clumps of matter (the halos) with some
density profile. The large-scale clustering of the mass in the linear (and
weakly non-linear) regime is given by that of the correlation between halos,
which should trace the linear (and weakly non-linear) perturbation theory
predictions, while the small-scale clustering of the mass arises from a
convolution of the halo density profile with itself. Thus, in this model the
results from weak-lensing magnification
 in right panel of Fig.\ref{x2c80sdss} mostly traces the halo
profiles.  

Smith \etal (2002) present new versions of the universal non-linear fitting
formulae $f_{NL}$ that incorporate these new ideas. The new formulae seems to
perform quite well against the Virgo N-body simulations. For the $\Lambda$CDM
model, the comparison involves scales up to $k \simeq 20$ h/Mpc, which
corresponds to $R \simeq \pi/k \simeq 0.16$ Mpc/h, comparable to the ones of
interest here.  We will therefore use this new formulae for $f_{NL}$ (as
implemented in the software made publicly available in Smith \etal 2002), to
compare to our results. As a first test, left panel of Fig.\ref{x2c80sdss}
shows the Smith \etal (2002) non-linear prediction (dashed-line) against our
N-body results from $\Lambda$CDM simulation (opened circles).  The agreement
is excellent and does not involve any additional parameter fitting. It should
be noted that the differences with previous fitting formulaes (eg Peacok \&
Dodds 1996) in this particular case are small. This is important as it means
that $\Lambda$CDM predictions for galaxy-QSO cross-correlations on non-linear
scales in the literature are correct and we need to look elsewhere for the
origin of the discrepancy.


As can be seen in the right panel of Fig.\ref{x2c80sdss} the weak-lensing
magnification predictions show a very steep non-linear variance, $\gamma
\simeq 2$, which in the context of stable model or direct simulation (see
\S\ref{sec:stable} above) can be interpreted as indication of a steep linear
spectrum $n_{eff} \simeq 0.9-1.0$.  This steep linear spectrum can be obtained
from the Smith \etal (2002) formulae by using a large effective value of the
shape parameter $\Gamma$. For example, $\Gamma=200$  ($\sigma_8=1$,
$\Omega_m=0.3$ and $\Omega_\Lambda= 0.7$) produces\footnote{This large
fiducial value for $\Gamma$ illustrates how different is the measured slope
from the expectations in the standard CDM picture.} $n_{eff}=0.96$.  This
prediction is shown as a dotted-line in left panel of Fig.\ref{x2c80sdss}.
This illustrates the point that we need a large positive slope to fit the
data.  Note that, once we fix the slope (from the same measurements) the only
free parameter is $\sigma_8$, which we take to be $\sigma_8=1$ for consistency
with power spectrum measurements (eg Gazta\~naga \& Baugh 1998, 
Tegmark \& Zaldarriaga 2002 and
references therein) and with the higher order correlations (see Bernardeau
\etal 2002 and references therein).

This means that the shape and amplitude for $\xibar_{NL}$ that we
find from weak-lensing are compatible with the general picture of non-linear
evolution of clustering and with the $\sigma_8=1$ normalization. Under this
assumptions the only difference with the {\it standard $\Lambda$CDM picture}
is that the linear spectrum on small (non-linear) scales has to be steeper. If
we match this with the {\it standard $\Lambda$CDM} $P(k)$ on larger scales,
this requires $P(k)$ to turn around and increase (rather than decrease) on
galactic scales.  As these scales are not strongly constraint by other
observations (see below), the contradiction can be resolved in this way.

\subsection{$\Omega_m$ dependence and Skewness}

Other alternatives to the above considerations on non-linear modeling
are to change the value of $\Omega_m$ and  
$\Omega_\Lambda$ but keeping $n_{eff} \simeq -1.5$ (eg with a low
 shape parameter). 
 
Open models with $\Omega_m \simeq 0.01$ have more power on small scales, just
as we need to explain the data.  This does not seem to work for two reasons.
First, because if we lower $\Omega_m $ the weak lensing magnification signal
becomes smaller which translates into a higher value of the recovered mass
$\xibar_{NL}$. Second, because the non-linear slope $\gamma$ is too low
compare to the weak-lensing reconstruction.
 
As mentioned above (end of \S6.4) closed models with $\Omega_m \simeq 1$
predict a three times stronger weak lensing magnification signal. This
translates into a nine times smaller reconstructed variance and three times
larger biasing, which is still not able to reconcile the discrepancy with the
standard predictions for the amplitude of non-linear matter fluctuations.

Bernardeau et al (1997) predicted $S_3 \simeq - 40 \Omega_m^{-0.8}
\zbar_{QSO}~^{-1.35}$ for the convergence skewness $S_3$ in the  $\Lambda$CDM model.
One should roughly expect  $S_3' \simeq -S_3$, so that
a direct comparison with Eq.[\ref{s3}] using our mean
$\zbar_{QSO} = 1.67$ yields 
$
\Omega_m \simeq 1.08 ~^{-0.31}_{+0.65}
$.
This is high and about 3 sigma out of 
the currently favored value of  $\Omega_m \simeq 0.2-0.3$ 
(eg Tegmark \& Zaldarriaga 2002 and references therein).
Note nevertheless that this is just a very rough comparison,
both because we need to take into account non-linear effects (see
Hui 1999) and because of the differences with pseudo-skewness. 
A direct comparison can be made to Fig.3 in 
M\'enard, Bartelmann \&  Mellier (2002), which corresponds to  $z_Q \simeq 1.5$.
Again here, the values for $S'_3$ in Eq.[\ref{s3}] favor the $\Omega_m \simeq 1$
and seem about 3-sigma away from the  $\Omega_m \simeq 0.3$ value. The 
$S_3$ predictions
are insensitive to the linear (scale dependence) bias and the amplitude and shape
of the linear mass spectrum. But note that they are quite sensitive to 
linear spectrum $n_{eff}$ of fluctuations on non-linear scales as they relay on
the  so-called saturation value in hyper-extended perturbation theory
(Scoccimarro \& Frieman 1999). If we change the value from $n_{eff} \simeq -1.5$
to $n_{eff} \simeq 0.9$, as suggested by the non-linear slope $\gamma$ (see
\S\ref{sec:stable} above) then we get a value for $S_3$ which is
about $5$ times smaller. We thus have $S_3 \simeq 8 \Omega_m^{-0.8}
\zbar_{QSO}~^{-1.35}$ which now yields:
\beq
\Omega_m \simeq 0.15 ~^{-0.05}_{+0.08}
\label{omegam}
\eeq
in better agreement with the  $\Lambda$CDM cosmology.

\subsection{Comparison with other results}

As mentioned in the introduction, previous measurements of galaxy-QSO
associations typically find evidence for weak lensing magnification.  In fact,
our results seem to agree well with the ones presented in Benitez \etal (2001
and references therein) which also seem at odds with current expectation
(see Bartelmann \& Schneider 2001).

Are these  results in contradiction with other measurements of the mass or of
the bias? On these scales we have measurements of the spectrum from Ly-alpha
forest (Croft \etal 2002), but they are mostly sensitive to the linear regime, as they
correspond to higher redshifts.  Weak shear lensing results (Hoekstra \etal 2002)
seem to agree well with the {\it standard  $\Lambda$CDM picture} 
above, and therefore seems at odds
with the results presented here. In particular, compare $b/r \simeq 1$ in
Fig.6 in Hoekstra \etal 2002 with our $b/r \simeq 0.1$ in Fig.\ref{w2bover},
which are both obtained under the same definitions of biasing parameters. Note
nevertheless that our analysis is probing slightly smaller scales, and more
importantly note that we do not assume any specific model for the underlaying
matter fluctuations.  If the  value of $\xi_{NL}$ is indeed as large as indicated by
our analysis one would also expect a stronger aperture mass variance, but
given the number of parameters involved in the comparison of data with theory,
some other interpretations  may still be possible (eg Croft \& Metzler 2000,
Mackey, White \& Kamionkowski 2002).  More work is needed to
understand this apparent discrepancy.
The values of $S_3$
found by Bernardeau, Van Waerbeke \& Mellier (2002) from the VIRMOS-Descart data
should also be compared 
 with the pseudo-skewness we find in Fig.\ref{s3edr}. 

\section{Conclusion}

Our results for a strong galaxy-QSO cross-correlation passed several test
presented in \S3 and \S4 and Fig.\ref{w2mz}-\ref{w2icut}.  In particular note
right panel in Figure \ref{w2icut} which shows how the signal disappears for
faint QSOs, as expected if produced by weak lensing magnification.  Also note
in top-left panel of Fig.\ref{w2icut} that the brighter the QSO sample the
larger the amplitude of the positive detection.   

Strong galaxy-QSO or cluster-QSO lensing magnification can also contribute to
the positive cross-correlation. Multiple QSOs images or just strong
magnification by chance alignment will typically give positive correlations,
in addition to the magnification bias effect. The key issue here is the
probability of strong lensing, which for a fixed QSO redshift increases with
the number density of QSOs.  Thus, one would expect this effect to be stronger
for the fainter QSOs, as the mean density is larger and the redshift
distribution is very similar.  This is contrary to what we find in
Fig.\ref{w2icut}.  Moreover, Benitez \etal 2001 argue that for strong lensing
$w_{GQ} \simeq f_S w_{GG}$, where $f_S$ is the fraction of QSOs in a sample
that are strongly lensed.  In Fig. \ref{w2gqfit} we find $w_{GQ} \simeq w_{GG}$
on the smallest scale for $\zbar =0.35$. This requires $f_S \simeq 1$, which
is unrealistic given the small size of the Eisntein radius of galaxies or clusters.
 Non-linear corrections to the weak-lensing approximation
are typically small for the variance (see M\'enard \etal 2002). Thus, weak
lensing magnification seems the most plausible explanation for the detected
galaxy-QSO correlation (for other interpretations see Burbidge \etal 1990).

Fig. \ref{w2gqfit} and Fig.\ref{x2c80sdss} summarized the main results
presented in this paper, which seems at odds with the standard  $\Lambda$CDM model, both
because the large amplitude and the steep slope of the recovered variance on
non-linear scales.  Note that this broad interpretation is independent of the
details in our modeling to recover the variance. 
As can be seen in Fig. \ref{w2gqfit} we find $w_{GQ} \simeq w_{GG}$
on the smallest scale for $\zbar =0.35$. This alone, indicates
that the bias factor $b$ has to be quite small, $b<<1$, as weak lensing is typically
less efficient that unity. In \S6 we argue that a possible explanation for this
strong correlation is that 
the effective linear spectral slope, $P(k) \sim k^{n_{eff}}$, on small scales is
 $n_{eff} \simeq 1$,  which reproduces the measured slope $\gamma \simeq 2$
in agreement with Eq.[\ref{stable}]. This is much steeper than in  $\Lambda$CDM, 
where  $n_{eff} \simeq -1.5$ and $\gamma \simeq 1.3$. 
  Within currently accepted cosmological
parameters ( $\sigma_8 \simeq 1$, $\Omega_m \simeq 0.3$ and $\Omega_\Lambda
\simeq 0.7$) adopting $n_{eff} \simeq 1$ can explain at the same time the
slope and amplitude of the recovered variance in Fig.\ref{x2c80sdss} and the 
low skewness in Fig.\ref{s3edr}. In the  $\Lambda$CDM cosmology this corresponds
to a bump or step on the power spectrum $P(k)$ on galatic scales.
A physical interpretation for this feature 
can be found in the early universe, where a bump or step in the primordial sprectum 
is expected (see Barriga \etal 2001). 
If such a primordial feature is really present,
it would have important implications for models
of galaxy formation. Our weak lensing analysis requieres
the bias $b$ to be a strong function of scale decreasing from $b \simeq 1$ 
at large scales to $b=0.1$ at galactic scales, thus hiding the presence
of such primordial bump. 
Note nevertheless, that if biasing is more complicated 
than modeled here (eg Eq.\ref{bgamma_b}), 
our quantitative interpretation could be different in detail (see the Appendix
for a discussion).

This is a preliminary analysis from a small fraction of the SDSS early
commissioning data. To explain the detected excess galaxy-QSO correlation of
$\sim 10\%$ with correlated photometric errors, such us flat-fielding or
scatter light, we would need very significant photometric variations: RMS of
at least $\Delta i' \simeq 0.3$ on arc-minute scales\footnote{This assumes unity
slopes for the number counts $N(m>i')$, realistic slopes require even larger
photometric errors.}, which is larger than the nominal calibration uncertainties 
of  0.03 magnitudes (see Stoughton et al 2002).
It is improbable that such large photometric errors are
present in the EDR/SDSS as they would have already been detected as an excess
galaxy-galaxy correlation when compared to other surveys (eg see Gazta\~naga
2002a, 2002b, Connolly \etal 2002, Scranton \etal 2002). Nevertheless, 
the full SDSS catalog should be able to
do a much better analysis of systematics that the one presented here and
confirm or refute our findings with a high significance.  A larger piece of
the SDSS would allow a comparisons to larger scales, which is missing in our
analysis because of the narrow width of the EDR strips. On larger scales, the
weak-lensing magnification signal can be more directly compared to other
indications of mass clustering, which will be an essential test for cosmology
and for the weak lensing magnification nature of these cross-correlations.

\section*{Acknowledgments}

I would like to thank R.Scranton and Josh Frieman for useful discussions
and comments on the SDSS/EDR data. I also like to thank
David Hughes, Marc Manera and Roman Scoccimarro
 for stimulating discussions.
I acknowledge support from supercomputing center at CEPBA and CESCA/C4, where
part of these calculations were done. I also acknowledge support from
and by grants from IEEC/CSIC and the Spanish Ministerio de Ciencia y
Tecnolog\'ia, project AYA2002-00850 and EC FEDER funding.

Funding for the creation and distribution of the SDSS Archive has
been provided by the Alfred P. Sloan Foundation, the Participating
Institutions, the National Aeronautics and Space Administration, the National
Science Foundation, the U.S. Department of Energy, the Japanese
Monbukagakusho, and the Max Planck Society. The SDSS Web site is
http://www.sdss.org/.



\appendix

\section{galaxy biasing}

We will consider two different mathematical approaches
to parameterize the effects of biasing, ie. how galaxy fluctuations
$\delta_G$ trace the underlaying mass distribution $\delta$.
More elaborated physical models are based on semi-analytical  galaxy
formation (see  eg Baugh, Cole \& Frenk 1996, Seljak 2000, Scoccimarro \etal
2001, Sheth \etal 2001, Cooray \& Sheth 2002)  
but the results are quite model dependent.
 
The first model consists in assuming that bias is linear, but non-local:
\beq
\delta_G(x) = \int~d^3 x' ~b(x-x')~\delta(x')
\eeq
Which in Fourier space is just:
\beq
\delta_G(k) = b(k)~ \delta(k)
\label{bk}
\eeq
This translates into a scale-dependence bias on the power-spectrum:
\beq
P_G(k) = b^2(k)~P(k)
\eeq
For a power-law correlation $P(k)$ and a power-law $b(k)$
we have that the galaxy variance is 
\beq
\xibar_G (R) = b^2_{0.2} ~\sigma_{0.2}^2 \left({0.2 Mpc/h\over{R}} \right)^{\gamma+2\gamma_b}
\label{bgamma_b}
\eeq
where $b_{0.2}$ and $\gamma_b$ characterize the amplitude and scale dependence
of the effective bias:
\beq
b(R)= b_{0.2} \left({0.2 \over{R}}\right)^{\gamma_b}.
\eeq  
If $\gamma_b=0$ this reduces to the standard linear biasing
model $\xibar_G(R) = b^2 ~\xibar(R)$.

In the second model we assume that bias is non-linear, but local:
\beq
\delta_G(r) = F[\delta(r)].
\eeq
For small fluctuations $\delta < 1$:
\beq
\delta_G \simeq b_1~ \delta
+ {b_2\over{2!}} ~ \delta^2 + {b_3\over{3!}} ~ \delta^3+ \Or[\delta^4]
\eeq
which gives rise to (see 
Fry \& Gazta\~naga 1993):
\beq
\xibar_G = b_1^2 ~\xibar +  c~\xibar^2 + d~\xibar^3 + \dots
\label{biasc}
\eeq
where $c= (S_3 b_2 b_1 + b_3 b_1 + 0.5 b_2^2)$ depends both on
the non-linear biasing parameters and $S_3$, the reduced skewness of the mass.
Thus, the local model formally gives
\beq
\xibar_G = {\cal F}[\xibar]
\eeq
which reduces to the standard linear biasing
model $\xi_G(r) = b^2 ~\xi(r)$ for $\xibar \rightarrow 0$. 
On non-linear scales we can write:
\beq
\xibar_G \simeq K \xibar^\phi ~~, \phi \equiv 
{d\log{\cal F}\over{d\log{\xibar}}}
\eeq
which reproduces Eq.[\ref{bgamma_b}] with  $\phi = 2 \gamma_b$
and $K = b^2_{0.2}/\sigma_{0.2}^{2(1+\gamma_b)}$. 

Thus we have shown that two quite different hypothesis about biasing
 drives  to a similar expression, ie Eq.[\ref{bgamma_b}], 
at least in the power-law limit. It is therefore plausible to assume that
a more generic non-local and non-linear biasing could also be cast
with such a parameterization. We therefore adopt Eq.[\ref{bgamma_b}]
 on the assumption that a power-law should be a good approximation for the
limited range of scales we consider here. We will be able to somehow
test this assumption with the data.

\subsection{Stochastic bias}

The stochasticity of the galaxy-mass
relation can also contribute to the observed galaxy clustering 
 (see Pen 1998, Scherrer \& Weinberg 1998, Tegmark \& Peebles 1998,
Dekel \& Lahav 1999, Matsubara 1999, Bernardeau \etal 2002, and references therein),
so that the relation between
$\delta_G$ and $\delta$ is not deterministic but rather stochastic,
\begin{equation}
\delta_G(r) = F[\delta(r)] + \epsilon_\delta(r)
\label{eulocbias2}
\end{equation}
\noindent where the random field $\epsilon_\delta$ denotes the
scatter in the biasing relation  at a given $\delta$ due to the
fact that $\delta(r)$ does not completely determine
$\delta_G(r)$. Under the assumption that the scatter is
local, in the sense that the correlation functions of
$\epsilon_\delta(r)$ vanish sufficiently fast at large
separations (i.e. faster than the correlations in the density field),
the deterministic bias results hold for the two-point correlation
function in the large-scale limit ( Scherrer \& Weinberg 1998).

This stochasticity is usually  characterized by a parameter $r$
defined as:
\beq
r \equiv {<\delta \delta_G> \over{b <\delta \delta>}} ~~~ {\rm with} ~~~
b^2  \equiv {<\delta_G \delta_G> \over{ <\delta \delta>}}
\eeq
which implies:
\beq
{r\over{b}} \equiv  {<\delta \delta_G> \over{<\delta_G \delta_G>}}
\label{roverb}
\eeq

Note that $r$ and $b$ are mean quantities and both contain information on the
scale independence and the stochastic $\epsilon$.  This provides a different
parameterization of biasing than the one given in the above subsection and has
been used by Hoekstra \etal (2002) to characterize the comparison of the
weak-lensing and galaxy correlations. To provide a comparison we also present
in Fig.\ref{w2bover} results for $b/r$ define in such a way, but note that the
meaning of $b$ here is quite different from the one in Eq.[\ref{bk}] (see
Dekel \& Lahav 1999). Note that the cross-correlation coefficient $|r| \le 1$, 
which introduces important constraints: eg $b/r <0.1 $ implies $b<0.1$.

The parameterization presented in the above subsection in terms of $b_{0.2}$
 and $\gamma_b$, is not necessarily in contradiction with the idea that the
 stochasticity could play an important role in biasing. We choose $b_{0.2}$
 and $\gamma_b$ for simplicity and to keep the number of parameters to a minimum. 
At each scale we can recover from our analysis the above stochastic modeling
by simply replacing our constraints for $b_{0.2}$ by constraints $b/r$ and 
our constraints for $\sigma_{0.2}$
by constraints on $r \sigma_{0.2}$. As $|r|<1$ we have that  
$\sigma_{0.2} > r \sigma_{0.2}$ which shows that our prediction
for $\sigma_{0.2}$, assuming $r=1$, is a lower bound.
Thus the qualitative conclusion in \S 7, that $\sigma_{0.2}$ is
too large as compare to $\Lambda$CDM models, is unchange by
this modeling of bias. 
Further work would be needed to disentangle  more complicated
biasing schemes.


\begin{thebibliography}{99}


\def\refe {\par \hangindent=.7cm \hangafter=1 \noindent}
\def\apj {ApJ}
\def\apl {ApJL}
\def\na {Nature,}
\def\aap {A\&A}
\def\apjs{ApJS}
\def\mn {MNRAS}

\bibitem[]{} Barriga, J., Gazta\~naga, E., Santos, M.G., Subir, S.
2001, MNRAS 324, 977



\bibitem[Bartelmann \& Schneider 1993]{1993A&A...271..421B} Bartelmann, M. 
\& Schneider, P. 1993b, A\&A, 271, 421 

\bibitem[]{1878}Bartelmann, M., \& Schneider, P. 1994, A\&A, 284, 1

\bibitem[]{Conn} Bartelmann, M., Schneider, P.,  
Phys. Rep., 2001, 340, 291

\bibitem[]{1883}Bartsch, A., Schneider, P. \& Bartelmann, M., 1997, A\&A, 319, 375

\bibitem[]{1885}Baugh, C.M., Cole, S., Frenk, C.S., 1996, MNRAS 283, 1361 

\bibitem[]{1887}Baugh, C.M., Gazta\~naga, E., 1996, MNRAS 280 L37

\bibitem[]{1889}Ben\'{\i}tez, N., \& Mart\'{\i}nez-Gonz\'{a}lez, 1995, ApJL 339, 53

\bibitem[]{1891}Ben\'{\i}tez, N., \& Mart\'{\i}nez-Gonz\'{a}lez, 1997,  ApJ, 477, 27

\bibitem[]{1893}Ben\'{\i}tez, N., Mart\'{\i}nez-Gonz\'{a}lez, E., Gonz\'{a}lez-Serrano,
J.I., \& Cay\'{o}n L. 1995, AJ, 109, 935

\bibitem[]{1896}Ben\'\i tez, N., Mart\'\i nez-Gonz\'alez, E. \& 
Mart\'\i n-Mirones, J. M. 1997, A\&A, 321, L1 


\bibitem[ref]{1902}Benitez, N., Sanz, J.L., Martinez-Gonzalez, E., 2001, MNRAS 320, 241





\bibitem[ref]{1910} Bernardeau, F., Van Waerbeke, 
  Mellier, Y. 1997, A\&A 322, 1

\bibitem[ref]{1913} Bernardeau, F., Van Waerbeke, 
  Mellier, Y. 2002, A\&A 389, L28

\bibitem[Bernardeau et al 2001]{1916}Bernardeau, F., Colombi., S, Gazta\~naga,
  E., Scoccimarro, R., 2002, Phys. Rep., 367, 1

\bibitem[ref]{1919}Blanton, and the SDSS collaboration, 2002 
2001, A.J., 121, 2358



\bibitem[ref]{ref} Burbidge, G. \etal, 1990, ApJS 74, 675

\bibitem[ref]{1924}Colombi, S., Szapudi, I., Szalay, A.S., 1998,  \mn 296, 253


\bibitem[ref]{1927}Connolly and the SDSS collaboration, 2002
astro-ph/0107417

\bibitem[ref]{1930}Cooray, A, Sheth, R., 
Phys. Rep, in press, astro-ph/0206508

\bibitem[ref]{1933}Croft, R.A.C. \etal 2002, ApJ in press, astro-ph/0012324

\bibitem[ref]{1935}Croft, R.A.C. \& Metzler, C.A.,  2000, 
ApJ 545, 561

\bibitem[ref]{1938}Croom, S.M., Shanks, T, 1999, \mn 307, L17

\bibitem[ref]{1940}Dekel, A., Lahav, O., 1999, ApJ 520, 24

\bibitem[ref]{1942}Dodelson, S., and the SDSS collaboration, 2002  
ApJ 572, 140

\bibitem[ref]{1945}Drinkwater, M. J., Webster, R. L., Thomas, P. A. \& Millar, E. 1992, 
Proceedings of the Astronomical Society of Australia, 10, 8 


\bibitem[]{1952} Frieman, J.A., Gazta\~naga, E., 1994, \apj 425, 392 

\bibitem[]{1954} Frieman, J.A., Gazta\~naga, E., 1999, \apj  521, L83



\bibitem[ref]{1959}Fry, J. N. 1994, PRL 73, 215

\bibitem[ref]{1961}Fry, J. N. \& Peebles, P.J.E., 1980, ApJ, 238, 785

\bibitem[ref]{1963}Fry, J. N. \& Gazta\~naga, E. 1993, ApJ, 413, 447

\bibitem[Fugmann 1988]{1988A&A...204...73F} Fugmann, W. 1988, A\&A, 204, 73 


\bibitem[ref]{1969}Fukugita, M., \etal 1996, AJ 111, 1748



\bibitem[Gazta\~naga 1994]{gaz94} Gazta\~naga, E., 1994, \mn
 268, 913 

\bibitem[]{1978}Gazta\~{n}aga, E., 1995, ApJ, 454, 561

\bibitem[]{gaz98} Gazta\~naga, E., \& Baugh, C.M. 1998, \mn, 294, 229 

\bibitem[]{1983}Gazta\~naga, E., 2002a, \mn 333, L21

\bibitem[]{1985}Gazta\~naga, E., 2002b, ApJ 580, 144
 
\bibitem[]{1987}Gazta\~naga, E., \& Bernardeau, F 1998, A\&A
331, 829 

\bibitem[]{1990}Guimar\~aes, A.C.C., van de Bruck, C.
Brandenberger, R., 2001, MNRAS, 325, 278 



\bibitem[ref]{1995}Hamilton, A.J.S., Kumar, P., Lu, E., Matthews, A., 1991, ApJ Lett., 374, L1

\bibitem[Hammer \& Le Fevre 1990]{1990ApJ...357...38H} Hammer, F. \& Le 
Fevre, O. 1990, ApJ, 357, 38 

\bibitem[]{2003_a}Hintzen, P., Romanishin, W., \& Vald\'{e}s, F., 1991, ApJ, 371, 49


\bibitem[ref]{2003_b}Hoekstra, H., Van Waerbeke, L., Gladders, M.D.,
  Mellier, Y., Yee, H.K.C.,  2002, Ap.J in press, astro-ph/0206103

\bibitem[]{2009}Hui, L.\& Gazta\~naga, E., 1999, \apj, 519, 622

\bibitem[]{2011}Hui, L., 1999, ApJL, 519, L9

\bibitem[]{2013}Kiss Cs, \etal 2002, astro-ph/0212094

\bibitem[Jenkins \etal 1998]{1998ApJ...499...20J} Jenkins, A., \etal 
1998, ApJ, 499, 20 


\bibitem[]{2019}Jain, B., Mo, H.J., White, S.D.M., 1995, MNRAS 276, L25

\bibitem[ref]{2018}Mackey, J, White, M., Kamionkowski M, 2002, astro-ph/0106364

\bibitem[Maddox 2000]{2020}Maddox, S.J., Efstathiou, G., Sutherland, W.J. \& Loveday, J., 1990,
\mn, 242, 43P

\bibitem[ref]{2023}Matsubara, T, 1999, ApJ 525, 543

\bibitem[ref]{2025}M\'enard, B., Bartelmann, M., 
  Mellier, Y,  2002, A\&A submitted, astro-ph/0208361

\bibitem[ref]{2028}M\'enard, B., Bartelmann, M.
 2002,  A\&A submitted astro-ph/0203163

\bibitem[ref]{2031}M\'enard, B., Hamana, T., Bartelmann, M., 
  Yoshida, N,  2002, A\&A submited, astro-ph/0210112

\bibitem[]{2040}Moessner, R. Jain, B, 1998, \mn 294, 291

\bibitem[]{2043}Navarro, J., Frenk, C., White, S.D.M, 1996, ApJ, 462, 563

\bibitem[]{2046}Norman, D. J. \& Impey, 
C. D. 1999, AJ, 118, 613 

\bibitem[]{2049}Norman, D.J. \& Williams, L.L.R. 2000, 
ApJ 119, 2060

\bibitem[]{2052}Peacock, J.A.,  \& Dodds S.J., 1994, MNRAS 267, 1020

\bibitem[]{2054}Peacock, J.A.,  \& Dodds S.J., 1996, MNRAS 280 L19


\bibitem[]{2057}Pen, U-L, 1998, ApJ 504, 601

\bibitem[Peebles 1980]{2051}Peebles, P.J.E., 1980, The
Large--Scale Structure of the Universe,
Princeton University Press, Princeton (LSS)


\bibitem[]{2067}Richards, G.T. \etal 2002, AJ 123, 2945


\bibitem[ref]{2062}Schneider, D.P. and the SDSS collaboration  
2002, AJ 123, 567

\bibitem[]{2073}Scoccimarro, R., Sheth, R.K., Hui, L., Jain, B. 2001a, \apj, 546, 20 

\bibitem[]{2076}Scoccimarro, R., Feldman, H., Fry, J. N.  \& Frieman, J.A., 2001b
\apj, 546, 652

\bibitem[]{2080}Scoccimarro, R., Frieman, J.A., 1999,  ApJ 520, 35, 44


\bibitem[]{2084}Scranton, R., and the SDSS collaboration
astro-ph/0107416, submitted to ApJ.

\bibitem[]{2087}Schneider, D.P.. \etal 2002, AJ 123, 567

\bibitem[]{2089}Scherrer, R.J., Weinberg, D.H., 1998, ApJ 504, 607


\bibitem[]{2092}Seljak, U.~; 2000, \mnras,
318, 203 

\bibitem[]{2092}Seldner M., Peebles, P.J.E.; 1979, ApJ 227, 30

\bibitem[Seitz \& Schneider 1995]{1995A&A...302....9S} Seitz, S. \& 
Schneider, P. 1995, A\&A, 302, 9 

\bibitem[ref]{Sheth} Sheth, R., Diaderio, A., Hui, L. Scoccimarro, R., 2001,
MNRAS, 326, 463

\bibitem[]{SzSz99} Smith, R.E., \etal  2002, astro-ph/0207664


\bibitem[]{2104}Stoughton C. and the SDSS collaboration, 2002
AJ 123, 485


\bibitem[]{2108}Szapudi, I. \& Gazta\~naga, E., 1998, \mn 300, 493

\bibitem[]{2110}Szapudi, I.,  Dalton, G.B. , Efstathiou, G.,  Szalay, A.S. 1995, ApJ 444, 520


\bibitem[]{2115} Szapudi, I., and the SDSS collaboration, 2002
ApJ 570, 75




\bibitem[]{2122}Tegmark, M., Hamilton, A.J.S., Strauss, M.A., Vogeley, M.S. \& Szalay, A.S. 
1998, \apj, 499, 555

\bibitem[]{2126}Tegmark, M., Zaldarriaga, M., 2002, astro-ph/0207047


\bibitem[]{2130}Tegmark, M.,  and the SDSS collaboration, 2002
ApJ 571, 191

\bibitem[]{2133}Tegmark, M., Peebles, P.J.E.,  1998
ApJL 500, 79


\bibitem[]{2137}Thomas, P.A., Webster, R.L., \& Drinkwater, M.J. 1994, MNRAS, 273, 1069

\bibitem[]{2139}Tyson, J.A., 1986, AJ, 92, 691




\bibitem[]{2148}Verde, L., \etal 2002, MNRAS in press

\bibitem[]{2150}Zehavi, I., and the SDSS collaboration, 2002
 ApJ 571, 172

\end{thebibliography}
\end{document}